\documentclass[11pt]{article}
\pdfoutput=1

\usepackage{amsmath}
\usepackage{amsfonts}
\usepackage{parskip}
\usepackage[margin=1in]{geometry}
\usepackage{float}
\usepackage{graphicx}
\usepackage{tikz}
\usepackage{tcolorbox}
\usepackage{subcaption}
\usepackage{algorithm}
\usepackage{algpseudocode}
\usepackage{xcolor}
\usepackage{tabularx}

\newcommand{\tensor}[1]{\boldsymbol{#1}}
\newcommand{\p}[2]{\frac{\partial #1}{\partial #2}}
\newcommand{\dq}[0]{\dot{\tensor{q}}}
\newcommand{\q}[0]{\tensor{q}}
\newcommand{\tdt}[0]{{t + \Delta t}}

\begin{document}

\title{Resolving Force Indeterminacy in Contact Dynamics Using Compatibility Conditions}
\author{
  Tyler Olsen, Ken Kamrin \\
  Department of Mechanical Engineering \\
  Massachusetts Institute of Technology
}
\maketitle

\section{Abstract}
Contact dynamics (CD) is a powerful method to solve the dynamics
of large systems of colliding rigid bodies.
CD can be computationally more efficient than classical penalty-based
discrete element methods (DEM) for simulating contact between stiff 
materials such as rock, glass, or engineering metals.
However, by idealizing bodies as perfectly rigid, contact forces
computed by CD can be non-unique due to indeterminacy in the contact
network, which is a common occurence in dense granular flows.
We propose a CD method that is designed to identify only the unique set of contact forces
that would be predicted by a soft particle method, such as DEM, in the limit of large stiffness.
The method involves applying an elastic compatibility condition
to the contact forces, which maintains no-penetration constraints but filters out force distributions that could not have arisen from stiff elastic contacts.
The method can be used as a post-processing step that could be
integrated into existing CD codes with minimal effort.
We demonstrate its efficacy in a variety of indeterminate problems,
including some involving multiple materials, non-spherical shapes, and nonlinear contact
constitutive laws.

\section{Introduction}
There are, broadly, two major classes of methods for discrete
simulations of granular materials.
The first class of methods is called the Discrete Element Method (DEM),
and its use for simulating granular flows was pioneered by
Cundall and Strack \cite{Cundall1979}.
In these methods, the equations of motion are numerically integrated
forward time, typically by computationally efficient explicit
timestepping algorithms \cite{Kruggel2008}.
Contact forces are computed by penalizing the overlap of adjacent bodies,
as an approximation to the forces generated by elastic deformation
of contacting bodies.
The model can be extended to include viscous damping effects, to
dissipate energy upon collision, and to include friction
\cite{Pao1955,Brilliantov1996,Zhang2005,Zhang2017}.
Since contact models are often derived from solutions to
related continuum elasticity boundary value problems for contacting bodies \cite{Hertz1882},
the contact forces of DEM take advantage of additional physics at the grain level 
that the rigid-body methods described below do not address.
For this reason, we consider well-resolved DEM to provide
physically valid answers (for a given contact model) and will assess
the accuracy of other methods using DEM as a baseline.
A major limitation of DEM methods, however, is their reliance on
explicit time integration schemes, which require that the time step
be sufficiently small to propagate an accoustic wave through the
assembly of bodies in contact in order to guarantee stability.
This can be an extremely restrictive time step, particularly when
the bodies are made up of geological or engineering materials such
as rock or steel.
The consequence of this time step restriction is that the user has
no choice but to resolve each and every contact event in fine detail,
even if they are only interested in the behavior of the system on
much larger time scales.
This timestep restriction can render the methods too computationally
expensive to be used for simulations that span practical lengths
of physical time when grains are stiff.
For a full discussion of time integration schemes for discrete element
methods, we refer you to \cite{Kruggel2008}.

The second class of methods, developed by the study of nonsmooth
mechanics, makes the simplifying assumption of perfectly rigid
bodies.
These methods are grouped under the term Contact Dynamics (CD), and
were pioneered by J.J. Moreau \cite{Moreau1977,Moreau1987,Moreau1993}
with significant contributions to the problem formulation and solution
algorithms in more recent years
\cite{Anitescu2006,Radjai2009,Smith2012,Krabbenhoft2012,Mazhar2015,Negrut2018}.
Assuming perfect rigidity has two primary impacts on the
method, distinguishing it from DEM methods.
First, the wave speed becomes infinite, which requires the use
of implicit time integration.
This allows large, stable timesteps, and long-term behavior
of rigid body systems can be studied, often at a great computational
savings compared to a faithful representation of the system in a 
DEM method \cite{Madsen2007}.
Second, forces can no longer be computed directly from particle
overlaps, because zero-penetration is enforced as a hard
constraint.
Rather, forces are typically computed as the solution of a
complementarity problem which enforces zero-penetration at contact.
This will be described in detail in the following sections
of this text.

The primary advantage to using CD rather than a penalty-based discrete
element method is the size of the time step that is attainable.
The implicit integration scheme allows very large time steps,
though each time step requires the solution of a large-scale
complementarity problem.
The tradeoff is that each CD time step requires several orders of
magnitude more effort than a DEM time step of equal system size.
Taking the stability tradeoff and computational effort tradeoff into
consideration, it is often computationally cheaper to simulate
large systems of quasi-rigid bodies using CD rather than DEM.
A CPU-time comparison between complemetarity-based and penalty-based
simulation methods has been carried out by
the SBEL group at the University of Wisconsin \cite{Madsen2007},
though there has been significant improvement in the efficiency
of CD solvers since that time \cite{Anitescu2006,Morales2008,Mazhar2015}.
Another paper \cite{Brendel2004} provides a phase diagram showing
the relationship between particle stiffness, system size, and
solution time.
It, too, confirms the result that there is always a critical stiffness
at which CD is more efficient than DEM.

A disadvantage of current CD formulations, however, is that the contact forces
computed to enforce the no-penetration constraint between bodies
are, in general, not unique.
The non-uniqueness of contact forces will be explored in detail in subsequent sections.
This brings into question the validity of complementarity-based simulation
approaches, especially when friction is considered.
Penalty methods do not suffer from this uniqueness problem.
Resolving the non-unique contact forces of CD is the main focus
of this work, and we present a method that ensures the CD algorithm only outputs forces
that could have arisen as the solution of a penalty-based DEM model
in the uniform limit of stiffness scale going to infinity.

It should be noted that CD does not necessarily require
bodies to be perfectly rigid.
A relatively recent work \cite{Krabbenhoft2012} develops
CD with finite stiffness elasticity, which lets particles
overlap as in DEM, but the time integration is carried out
implicitly.
However, introducing finite elasticity seems to
re-introduce a finite acoustic wave speed, which reintroduces
restrictions on the time step required to accurately resolve
collision processes, even if the numerical integration is stable for larger time steps.
As in DEM, introducing finite elasticity ``regularizes''
the problem to obtain unique force distributions, but the authors note that
the scheme still produces indeterminate forces in the limit of
perfect rigidity.
The present work addresses the issue of non-unique forces
while preserving the assumption of perfect rigidity, avoiding
the issue of introducing an acoustic wave speed entirely.
There have been efforts to quantify the degree and the effects
of indeterminacy in rigid body CD \cite{Radjai1996,Mcnamara2004},
however we are not aware of any previously proposed methodology to
resolve the indeterminacy without leaving the rigid-particle regime.
We are aware that as grains get stiffer, more physical uncertainty can arise in the force distribution,
since small surface or material perturbations have a larger effect on the outcome.
Even with this reality in mind, our  approach is still useful for identifying
the ideal force distributions under a given best-guess at contact physics,
which could serve as a characteristic or average field in light of the uncertainty.

\subsection{Compatibility Conditions}\label{compat_ex}
We approach the non-uniqueness of forces by seeking to impose
a ``compatibility condition'' on the force network that would
guarantee that the forces could have come from elastic particles in the limit
of infinite stiffness.
We are inspired by the use of compatibility conditions in the
solution of continuum PDE boundary value problems.
One example is (source-free) steady heat transfer, stated as continuity of heat flux, $\tensor{Q}$,
\begin{equation}
 \nabla \cdot \tensor{Q} = 0
  \label{eq::HeatEquation}
\end{equation}
subject to flux boundary conditions.
On its own, there are multiple solutions to the above equation.  
This indeterminacy is resolved by requiring the heat flux be related to a
temperature field by Fourier's law
\begin{equation}
  \tensor{Q} = -k\nabla T
  \label{eq::FouriersLaw}
\end{equation}
For simplicity, we assume a constant, scalar thermal conductivity $k$.
Traditionally, this system is solved by rewriting \eqref{eq::HeatEquation}
in terms of the underlying temperature field using \eqref{eq::FouriersLaw}
and solving for $T$, and then calculating $\tensor{Q}$ from $T$.
However, if one wants to solve the problem without resorting to calculating a temperature
field, a compatibility condition on the heat flux can be imposed by taking the curl of Fourier's law:
\begin{equation}
  \nabla \times \tensor{Q} = \tensor{0}.
\end{equation}
This condition ensures that a temperature field \textit{exists} from which
the heat flux field solving \eqref{eq::HeatEquation} is derived.
Combining this equation with (\ref{eq::HeatEquation}) allows one
to solve for $\tensor{Q}$ directly, provided flux boundary conditions.
This is true even in the limit as conductivity $k$ approaches infinity,
since $k$ factors out of the compatibility constraint,
thereby exiting the system of equations entirely.  Hence $\tensor{Q}$ can be found in this limit without
having to calculate the (vanishingly small) $T$ field first.

A similar application of compatibility arises in continuum elasticity.
For linearized elasticity, the equilibrium PDE is
\begin{equation}
  \nabla \cdot \tensor{\sigma} + \tensor{b} = \tensor{0}
  \label{eq::ElasticEquilibrium}
\end{equation}
for stress field $\tensor{\sigma}$ and body force $\tensor{b}$.
Like the previous example, there are multiple stress solutions to this system.
The indeterminacy is resolved by enforcing the constitutive equation
$\tensor{\sigma} = \tensor{\mathbb{C}}\tensor{\varepsilon}$, where $\mathbb{C}$
is the fourth-rank stiffness tensor of the material, and the strain $\tensor{\varepsilon}$
is computed from the displacement vector field $\tensor{u}$ using the strain-displacement relation
$\tensor{\varepsilon} = \frac{1}{2}\left(\nabla \tensor{u} + \nabla \tensor{u}^T\right)$.
The most common approach to solving boundary value problems is to
formulate \eqref{eq::ElasticEquilibrium} in terms of displacements,
solve for the displacement field, and compute stress and strain from displacement.
However, this is not the only way to solve for a stress field satisfying
\eqref{eq::ElasticEquilibrium} under traction boundary conditions.
A compatibility condition can be applied to the strain field instead to ensure
that an integrable displacement field exists.
The compatibility once again takes the form of a curl constraint \cite{Barber2002},
this time involving the curl of a second-order tensor.
\begin{equation}
  \nabla\times (\nabla \times \tensor{\varepsilon}) = \tensor{0}
  \label{eq::ElasticCompatibility}
\end{equation}
Moreover, a stress-only system emerges by restating the compatibility condition
using the constitutive relation,
\begin{equation}
  \nabla\times (\nabla \times (\tensor{\mathbb{C}}^{-1}\tensor{\sigma})) = \tensor{0}
  \label{eq::ElasticCompatibility2}
\end{equation}
Suppose the elastic stiffness tensor is decomposed as $\mathbb{C}=c\ \tilde{\mathbb{C}}$,
where $c$ is a stiffness scale and the relative stiffness $\tilde{\mathbb{C}}$
is order-one in size.
Then the infinite stiffness limit of $c\to\infty$ at fixed $\tilde{\mathbb{C}}$
can be captured by factoring out $c^{-1}$ from the above, which amounts to solving
(\ref{eq::ElasticEquilibrium}) simultaneous with
$ \nabla\times (\nabla \times (\tensor{\tilde{\mathbb{C}}}^{-1}\tensor{\sigma})) = \tensor{0}$.
Formulated in this way, the system of equations remains well-posed in
the limit of infinite elastic moduli, allowing limiting stress solutions
to be found directly without calculating a displacement field.
This approach is useful to compute stress distributions in stiff media,
whose displacement field would be almost imperceptibly small.

In a later section, we appeal to similar logic in our method for finding a
compatible set of contact forces in frictionless CD, ensuring existence of a 
microscopically small displacement field among contacting grains from which the forces 
could be derived, without ever solving for that underlying displacement field.

\section{Notation}
A system consists of $N_b$ bodies, and
quantities subscripted with a capital Roman letter ($A$)
indicate that the quantity is associated with a single
body with ID $A\in \{1 \dots N_b\}$.
Each body has a position and orientation, held in a
single generalized coordinate
$\q_A = [\tensor{r}_A^T, \; \epsilon_A^T]^T$,
with its time derivative
$\dq_A = [\dot{\tensor{r}}_A^T, \; \dot{\epsilon}_A^T]^T$ \cite{Negrut2018}.
The time derivative of the euler angles $\dot{\epsilon}_A$ are not easy to use
when posing equations of motion, however, so we instead use
a generalized velocity vector $\tensor{v}_A = [\dot{r}_A^T, \; \omega_A^T]^T$,
which can be related to $\dq_A$ using the velocity transformation matrix \cite{Haug1989},
\begin{equation}
  \dq_A = L_A(\q_A)\tensor{v}_A.
  \label{eq::velocityTransformation}
\end{equation}
$\tensor{v}_A$ and $\q_A$ have six components in 3D and
three components in 2D, corresponding to the free
translational and rotational degrees of freedom
in each case.
For the sake of clarity, we restrict ourselves
to discussing the 3D case in the remainder of the text,
but everything herein applies equally to the 2D
case as well.
As a matter of convention, spatial vector and tensor quantities on bodies,
such as generalized position $\tensor{q}_A$ and rotational inertia 
$\tensor{J}_A$, will be in boldface;
Scalars, global vectors, and global matrices will not.
The dimension of a vector or matrix will be either explicitly defined
or unambiguous from the surrounding context.
The global position and velocity vectors, $q,v \in \mathbb{R}^{6N_b}$,
indicated by the omission of a subscripted body ID,
are formed by the concatenation of all $\q_A$ and
$\tensor{v}_A$ vectors respectively.
\begin{equation}
  q = [\q_1^T \;\; \q_2^T \;\; ... \;\; \q_{N_b}^T]^T \in \mathbb{R}^{6N_b}
\end{equation}
All other global vectors of body-local quantities are formed from 
the concatenation of their single-body counterparts in the same way.

In addition to position and velocity information,
each body has a generalized mass matrix
$M_A \in \mathbb{R}^{6\times6}$, holding the
body mass $m_A$ and rotational inertia $\tensor{J}_A$
in such a way that the linear and angular momentum
of a body may be obtained with $M_A\tensor{v}_A$.
The rotational inertia tensor $\tensor{J}_A$ is defined with respect
to the body center of mass, and its components are
expressed with respect to a global, stationary
frame of reference.
Because the inertia is always stored with respect
the global frame of reference, the rotational components
of $M_A$ must (in general) be updated as the body rotates.
\begin{equation}
  M_A =
  \left[
    \begin{matrix}
      m_A & & & & & \\
      & m_A & & & & \\
      & & m_A & & & \\
      & & & J_{xx}^A & J_{xy}^A & J_{xz}^A \\
      & & & J_{yx}^A & J_{yy}^A & J_{yz}^A \\
      & & & J_{zx}^A & J_{zy}^A & J_{zz}^A \\
    \end{matrix}
    \right]
  \label{eq::MassMatrix}
\end{equation}
The global mass matrix $M \in \mathbb{R}^{6N_b \times 6N_b}$
is a block-diagonal matrix, with each $M_A$ positioned
as expected on the diagonal.

Besides bodies, the other system entity that must be considered
is a ``contact.''
A contact is simply a pair of bodies with the point of contact
and a unit normal vector of the contact.
We let $N_c$ be the number of contacts in our system.
At each contact ``$i$'', we are able to compute the signed distance between
the closest points of two associated (convex) bodies, denoted $\Phi_i$.
This quantity is the ``gap function,'' and is a vector-valued
function of the system position vector, $\Phi(q) \in \mathbb{R}^{N_c}$.

Importantly, $\Phi(q)$ is a function, and we are able to
compute its gradient with respect to the components of $q$.
In practice, we only consider contacts whose gap function
is less than some threshold (i.e. we only consider
contacts where the bodies are already or nearly in contact).
We consider the methods for computing $\Phi(q)$ and its
gradient an implementation detail, and will not address them
further here.
A similar ``gap function'' formalism is used in \cite{Negrut2018},
where more implementation details are provided.
An authoritative reference for the implementation of contact detection
systems is \cite{Ericson2004}.

\section{CD Formulation as Complementarity Problem}
In this section, we formulate CD for frictionless rigid-body contact.
We will restrict our attention to the frictionless case
for the remainder of the paper.

\subsection{Continuous-time Complementarity Condition}
Recall that in penalty methods, the force at a contact is computed
explicitly from the overlap of two bodies at a point.
In contact dynamics, however, bodies are idealized as perfectly rigid,
so it is no longer possible to specify an explicit constitutive law.
The forces, then, must be computed from some other condition.
In CD, contact forces serve to enforce the no-penetration constraint.
Furthermore, the force between two bodies
must be zero if the bodies are not in contact.
From these conditions, we can reproduce the ``Signorini conditions''
for a rigid contact
\begin{align}
  \Phi_i(q(t)) &> 0  \implies f_i(t) = 0 \nonumber \\
  f_i(t) &> 0 \implies \Phi_i(q(t)) = 0 \nonumber \\
  f_i(t) &\ge 0 \quad \text{no tension} \nonumber \\
  \Phi_i(q(t)) &\ge 0 \quad \text{no penetration}
  \label{eq::Signorini}
\end{align}
where $f_i(t)$ is the magnitude of the non-negative normal contact force
at contact ``$i$'' and $\Phi_i(q(t))$  is the gap function, as previously discussed.
These conditions establish a complementarity relationship between
the contact force and the gap function, $f_i(t) \Phi_i(q(t)) = 0$.
  
All together, these conditions can be encoded for all contacts in a
system with the statement
\begin{equation}
  0 \le f(t) \perp \Phi(q(t)) \ge 0
  \label{eq::Complementarity1}
\end{equation}
where $f\in\mathbb{R}^{N_c}$ is a vector containing the force at
all contacts, and $\Phi(q(t))$ is the full vector-valued gap function
at time $t$.
In \eqref{eq::Complementarity1}, the inequalities are understood
to apply component-wise to their respective vector operands.

\subsection{Equations of Motion and Time Integration}
The equations of motion for a body ``$A$'' can be written out using generalized
coordinates and velocities as
\begin{align}
  \dq_A &= L_A(\q_A)\tensor{v}_A \nonumber \\
  M_A \dot{\tensor{v}}_A &= \sum \tensor{F}_A
  \label{eq::EoMContinuous}
\end{align}
where $\sum \tensor{F}_A$ denotes the sum of forces acting on body $A$
including contact forces and externally-applied forces and moments (e.g. gravity).
Generalizing the equations of motion to the entire system is simply
a matter of removing the subscript $A$.
In \cite{Radjai2009}, the authors treat the system as a set of
measure differential equations due to the non-smooth nature
of rigid contacts.
However, we choose to retain the original ordinary differential equation
notation since (a) the rigidity assumption should be interpreted as the
limit of deformable bodies whose stiffness approaches $\infty$;
and (b) it does not affect the software implementation, since both
formulations yield the same system after discretization in time.

From here forward, we use the notation $[\cdot]^{t} = [\cdot](t)$ to refer to
a continuous-time quantity evaluated at time $t$.
We choose a semi-implicit Euler time integration scheme, as in
\cite{Negrut2018} and \cite{Radjai2009}:
\begin{align}
  q^{\tdt} - q^t &= \Delta t L(q^t)v^{\tdt} \nonumber \\
  M\left(v^{\tdt} - v^t\right) &= \Delta t F^{\tdt}
  \label{eq::EoMDiscrete}
\end{align}
where we drop the explicit $\sum$ from \eqref{eq::EoMContinuous} 
and understand $F$ to be the global vector of net forces on bodies.

Since this is an implicit time integration scheme, the forces
and velocities are computed at the end of a time step $[t, \tdt]$.
In practice, this means computing the contact forces at $\tdt$
and summing them onto the appropriate bodies.

\subsection{Discretized Complementarity Condition}
The time integration scheme above is used to discretize the complementarity
condition.
Like other formulations of CD \cite{Negrut2018},
we impose the complementarity condition at the end of the timestep in order to
ensure that the no-penetration constraint is satisfied at all steps of a simulation.
\begin{equation}
  0 \le f^{\tdt} \perp \Phi^{\tdt} \ge 0
  \label{eq::ComplementarityTimeDiscrete}
\end{equation}

At this stage, we approximate $\Phi^{\tdt}$ using a Taylor expansion
\begin{equation}
  \Phi^\tdt \approx \Phi^t + \Delta t Bv^{\tdt}
\end{equation}
where $B$ is a matrix that gives the change of the gap function for 
a set of infinitesimal displacements and rotations.
With a slight abuse of notation, we say that $B = \p{\Phi}{q}$.
The matrix $B \in \mathbb{R}^{N_c \times 6N_b}$, the gap function gradient,
has the property of mapping ``body'' quantities (e.g. velocity) to ``contact normal''
quantities (e.g. relative velocity in the contact-normal direction).
From a practical standpoint, we have to evaluate $B$ at the beginning of the time step
and we make the assumption that $B$ does not change substantially during the step.
This is a good assumption in a sufficiently-resolved simulation and is standard practice
in CD formulations \cite{Negrut2018}, \cite{Radjai2009}.

Thus, the complementarity condition can be re-written in terms of the known (beginning
of time step) gap function, and the end-of-timestep velocities.
\begin{equation}
  0 \le f^{\tdt} \perp \frac{\Phi^{t}}{\Delta t} + Bv^{\tdt} \ge 0
  \label{eq::ComplementarityTmp}
\end{equation}

Next, \eqref{eq::EoMDiscrete} may be substituted into \eqref{eq::ComplementarityTmp}
to obtain a complementarity condition purely in terms of the forces at the end of the
time step.
\begin{equation}
  0 \le f^\tdt \perp \frac{\Phi^{t}}{\Delta t}
  + B \left( v^t + \Delta t M^{-1}F^{\tdt}\right) \ge 0
  \label{eq::ComplementarityTmp2}
\end{equation}

The summation of forces can be carried out using an operator $H \in \mathbb{R}^{6N_b \times N_c}$
to accumulate contact forces onto bodies
\begin{equation}
  F^{contact} = H f
  \label{eq::SumForces}
\end{equation}
where $F^{contact} \in \mathbb{R}^{6N_b}$ is the global vector
containing $\tensor{F}^{contact}_A \in \mathbb{R}^{6}$,
the resultant force and moment of contacts on particle $A$,
for $A \in \{1, 2, ..., N_b\}$.
It can be shown by a virtual power argument that $H = B^T$, so the final form of the
complementarity condition is
\begin{equation}
  0 \le f^\tdt \perp \frac{\Phi^{t}}{\Delta t}
  + B \left( v^t + \Delta t M^{-1}\left(B^T f^{\tdt} + F^{ext}\right) \right) \ge 0
  \label{eq::ComplementarityDiscrete}
\end{equation}
where $F^{ext}$ is the force due to externally-applied loads (e.g. gravity)
and is given as data.

For the sake of brevity going forward, let matrix $N$ and vector $p$ be defined as
\begin{align}
  N &= \Delta t B M^{-1} B^T \nonumber\\
  p &= \frac{\Phi^{t}}{\Delta t} + Bv^t + \Delta t BM^{-1}F^{ext}
  \label{eq::NPAbbreviations}
\end{align}
which lets us rewrite \eqref{eq::ComplementarityDiscrete}
(dropping the implicit ``$\tdt$'' superscripts) as
\begin{equation}
  0 \le f \perp N f + p \ge 0
  \label{eq::ComplementarityFinal}
\end{equation}

The problem in \eqref{eq::ComplementarityFinal} is a canonical Linear Complementarity Problem (LCP), and
many solution methods exist to solve it efficiently. See \cite{Morales2008,Cottle2009}
and contained references for a comprehensive exploration of the topic.

\textbf{A brief note about restitution:}
The above formulation represents a perfectly-plastic model of rigid body contact.
If some restitution is desired, the formulation must be changed slightly to
account for this.
The (Newtonian) normal coefficient of restitution $e_n$ for two-body contact
is classically defined as $e_n = -\frac{\Delta v^{out}}{\Delta v^{in}}$,
with $\Delta v^{in (out)}$ as the signed relative velocity of two bodies
before (after) impact.
To account for restitution, complementarity is no longer
enforced at the ``gap'' level.
Rather, complementarity is enforced at the velocity level, and an implementation
must be careful not to consider any contacts that have not yet formed.
After some algebra, it can be shown that this is equivalent to replacing
the $\frac{\Phi^t}{\Delta t}$ term with $e_n B v^t$ in the expressions
\eqref{eq::ComplementarityTmp} through \eqref{eq::ComplementarityFinal}.
Other works such as \cite{Anitescu1997} assume the Poisson hypothesis for restitution,
wherein contact events are decomposed into ``compression'' and ``decompression''
phases.
However, this assumption requires the solution of two LCPs in each
time step, and is not considered in this work.
For the rest of the paper, we present the zero-restitution formulation,
but the above substitution can be made at any point to give non-zero restitution.

\section{CD Formulation as QP}
One popular method to solve the LCP in \eqref{eq::ComplementarityFinal}
is to re-formulate the problem as a quadratic program (QP) whose
solution also happens to solve the original LCP.
Consider the following QP:
\begin{align}
  \min_f \;\; &\frac{1}{2} f^T N f + f^T p \nonumber \\
  s.t. \;\; &f \ge 0
  \label{eq::CDQP}
\end{align}
with $N$ and $p$ defined as before in \eqref{eq::NPAbbreviations}.
It is worth noting that $N$ is symmetric and positive \textit{semi}-definite.
For a vector $f^*$ to be an optimal point of this QP, it must satisfy
the QP's Karush-Kuhn-Tucker (KKT) conditions---a set of
first-order necessary conditions in addition to the constraints of the
original problem for optimality \cite{Boyd2004}.
For \eqref{eq::CDQP}, the KKT conditions are
\begin{align}
  Nf^* + p - \lambda &= 0 \nonumber \\
  \lambda &\ge 0 \nonumber \\
  f^* &\ge 0 \nonumber \\
  f^{*T} \lambda &= 0
  \label{eq::CDKKT}
\end{align}
where $\lambda$ is a KKT multiplier (generalized Lagrange multiplier)
associated with the $f \ge 0$ constraint, which is complementary
to $f^*$.
From \eqref{eq::CDKKT}, we can see that since $\lambda = Nf + p$,
a solution of \eqref{eq::CDQP} will also satisfy the complementarity
formulation of the problem \eqref{eq::ComplementarityFinal}.

\subsection{Non-uniqueness of Forces}
From the QP formulation, we also have information about the uniqueness of
optimal solutions.
If $N$ is full-rank, then it will be positive-definite, \eqref{eq::CDQP}
is strictly convex, and the problem has a global and unique solution.
In this instance, there are no problems with this formulation of CD in calculating unique contact forces.
Classically, this situation occurs in isostatic packings, where force balance alone
determines all system forces.
However, if $N$ is rank-deficient, then (in general) there is a space 
of optimal solutions to the problem whose dimension is equal to the
dimension of the null-space of $N$ \cite{Boyd2004}.
This indeterminate case corresponds to packings that are more
highly-coordinated than isostatic. 

This situation is directly analogous to that of a statically indeterminate truss.
Recall that, much like the examples from Sec \ref{compat_ex}, unique solutions
to statically indeterminate truss problems are recovered by allowing joint
displacements, which are then used to relate truss member forces
to their corresponding elongations under an elasticity relation.
Relating system forces to a set of joint displacements poses
what amounts to additional system constraints, which filter out
all but one of the statically admissible solutions.
In the same way, the notion of contact stiffness and the requirement that contact 
forces arise from contact displacements resolves indeterminacy in particle systems.
This is true regardless of whether the system is static or dynamic.

Our goal in the upcoming sections is to show how this
requirement can be imposed as a force-only compatibility
constraint in the limit of stiff particles.
Unlike a standard truss, since we consider cohesionless
granular media it is crucial that the formulation be able
to admit contact separation in order to preclude contact tension.
Thus, we seek conditions on the contact forces that ensure
they could arise from a no-tension, compressibly stiff `truss'.

\section{Rigid Limit of Elastic Particles}
In this section we define the idea of a ``compatible'' set of contact forces
arising in the limit of infinitely stiff particles.

\subsection{Single Contact}

To begin, we examine a single visco-elastic contact whose force obeys a constitutive
law
\begin{equation}
  f = \left\{
    \begin{matrix}
    \hat{f}_e(c) + \gamma \hat{g}(c)\dot{c} &  \text{ if } c \ge 0\\
    0 & \text{ if } c < 0
    \end{matrix}
    \right.
  \label{eq::viscoelasticConstitutive}
\end{equation}
where $c$ is the overlap distance of particles at the contact,
$\dot{c}$ is the time rate of change of the
particle overlap distance, and $\gamma \hat{g}(c)$ is a viscous damping coefficient,
which in general can be a nonlinear, continuous function of $c$.
In this constitutive law, $\hat{f}_e(c)$ denotes the elastic constitutive law
(e.g. Hookean or Hertzian contacts), and is continuous for $c\in\mathbb{R}$.
By convention, we consider forces to be positive in compression.
We define a ``contact event'' as the interval in time ($t_c$) during which
$c \ge 0$.
With this in hand, the (scalar) impulse delivered to a particle by a contact can
be computed by integrating over time through a contact event.
\begin{equation}
  \int_{t_c} f\left(c(t),\dot{c}(t)\right) \,\mathrm{d}t  =
  \int_{t_c} \hat{f}_e(c(t)) \,\mathrm{d}t + \int_{t_c} \gamma\hat{g}(c(t)) \dot{c}(t)\,\mathrm{d}t 
  \label{eq::contactImpulse}
\end{equation}
By integrating over a complete contact event (i.e. $t\in (t_1, t_2)$
where $c(t_1)=0$, $c(t_2) = 0$, and $c(t) > 0$), we see that the
viscous term in \eqref{eq::contactImpulse} contributes exactly \textit{zero}
to the resulting impulse.
\begin{align*}
  \text{Viscous impulse} &= \int_{t_1}^{t_2} \gamma\hat{g}(c(t)) \dot{c}(t)\,\mathrm{d}t \\
  &= \int_{c(t_1)}^{c(t_2)} \gamma \hat{g}(c) \,\mathrm{d}c \\
  &= \int_{0}^{0} \gamma \hat{g}(c) \,\mathrm{d}c \\
  &= 0
\end{align*}

Another limit in which viscous impulse disappears is that of an
``enduring'' contact.
We denote a contact as ``enduring'' if, during some interval $t \in (t_1, t_2)$,
$c(t_1) = c(t_2) \ge 0$.
The viscous impulse is trivially zero in this case.\footnote{An
interesting observation, however, is that the impulse delivered by
the elastic part of the constitutive law \textit{is} impacted by changing
the damping behavior.
The effect of viscous damping appears to be, as expected, dissipating
energy during a collision by affecting the path $c(t)$.}

By limiting our scope to these cases where the viscous contribution to the impulse
is zero, it follows that the impulse delivered by a contact is simply the
time-averaged \textit{elastic} contact force.
However, by the mean value theorem, there exists some overlap $c^*$ such that
\begin{equation}
  \int_{t_1}^{t_2} \hat{f}_e(c(t)) \,\mathrm{d}t = \hat{f}_e(c^*)(t_2 - t_1)
  \label{eq::compatibleOverlapDefinition}
\end{equation}
This special overlap $c^*$ will be important when defining the forces from
a list of contacts being resolved simultaneously.

\subsection{Multiple Contacts CD}
One assumption made in CD is that during a time step $\Delta t$,
all contacts are resolved \textit{simultaneously}.
This assumption allows us to enforce the no-penetration condition at each
contact simulaneously, and thus to use the formulation of CD given in previous
sections. Additionally, in CD, we take large timesteps relative to the time
interval of a contact event,
so all contacts fall into one of the two previously-discussed limits
where either (a) a collision is fully resolved within a CD timestep $\Delta t$,
or (b) a contact is enduring.
In either case, there is zero net impulse delivered directly by the viscous
damping, so there must be a value $c^*$ at each contact from which we can
compute the contact impulse.

From these assumptions, we propose a method to
constrain CD to only admit force distributions that are ``compatible,''
i.e. force distributions that could be constructed from a set of
microscopic overlaps $c^*$ defined by a microscopic displacement
field of the particles.
This displacement is distinct from the particle motion given by
$q$ --- physically, it represents only the small part of the total
displacement responsible for creating elastic compression of grains.
As stiffness goes to infinity, this micro-displacement field vanishes,
but the limiting force distribution is unique.
Hence, our approach maintains the no-penetration conditions but appends
the compatibility constraint to ensure this limit is appropriately captured.

The compatibility condition amounts to requiring that the contact force
distribution must arise as the solution of a no-tension elastic
truss defined by the contact network between a collection of contacting bodies.
At a large but finite stiffness, each truss member $i$ has a value $c^*_i$
from which its force is computed, and the overlaps $c^*_i$ must be computed
from a single set of displacements:
\begin{equation}
  c^*_i = \hat{c}^*_i(u)
  \label{eq::compatibilityCondition}
\end{equation}
where $u \in \mathbb{R}^{6 N_b}$ is the global displacement vector of truss joints.
Since we are interested in a truss in the limit of small displacements,
we may safely restrict ourselves to the small overlap formula.
\begin{equation}
  c^* = -Bu
\end{equation}

\subsection{Zero-Tension Truss}

A no-tension truss subjected to arbitrary external loading
may be formulated in the following way (dropping $*$'s).
First define a constitutive law for the strain-energy
of a truss member:
\begin{equation}
  w_i(c_i) = \left\{
  \begin{matrix}
    w_{ei}(c_i) & \text{ if } c_i \ge 0 \\
    0 & \text{ if } c_i < 0
  \end{matrix}
  \right.
  \label{eq::Constitutive1}
\end{equation}
In \eqref{eq::Constitutive1}, $w_{ei}(c)$ is the constitutive law of
an elastic truss member and is defined for all $c$.
For example, linear truss members are defined by $w_{ei}(c_i) = \frac{1}{2}k_e c_i^2$.
Contact forces are related to the stored energy function \eqref{eq::Constitutive1}
by the subgradient
\begin{equation}
  f_i \in \partial w_i(c)
\end{equation}
which reduces to a simple derivative when $w_i$ is continuously differentiable.
\begin{equation}
  f_i = \left\{
  \begin{matrix}
    \frac{\mathrm{d}w_{ei}(c_i)}{\mathrm{d} c_i} & \text{ if } c_i \ge 0 \\
    &\\
    0 & \text{ if } c_i < 0
  \end{matrix}
  \right.
  \label{eq::Constitutive2}
\end{equation}

We consider a truss with zero imposed boundary displacement on degrees of freedom $\Gamma_D$
and arbitrary body forces $\tilde{F}$ on degrees of freedom $\Gamma_N$.
Formally, $\Gamma_D$ is the index set defined such that
$\Gamma_D = \{ j \;|\; u_j = 0, j\in \{1, 2, ..., 6N_b\} \;\}$.
The ``non-boundary'' index set $\Gamma_N$ then is $ \Gamma_N = \{j \; | \; j \notin \Gamma_D\}$.
With these definitions in hand, we apply the minimum potential energy
variational principle to the truss to find the equilibrium solution.

\begin{align}
  \min_{\tensor{u}, \tensor{c}}\;
  &\sum_{i=1}^{N_c} w_i(c_i) - \sum_{j \in \Gamma_N} \tilde{F}_j u_j \nonumber \\
  s.t. \; & c = -Bu \nonumber \\
  & u_j = 0 \;\; \text{ for } j \in \Gamma_D
  \label{eq::PrimalPE}
\end{align}

Following the formulation of a cable network in \cite{Kanno2011},
which is identical to this problem save for the sign convention,
\eqref{eq::PrimalPE} can be equivalently stated in terms of
the minimization of complementary energy.
In this form, the minimization is carried out over the contact forces $f_i$.

\begin{align}
  \min_{f} \;& \sum_{i=1}^{N_c} w^*_{ei}(f_i) \nonumber \\
  s.t. \;& (B^T f)_j = \tilde{F}_j \;\; \text{for } j \in \Gamma_N \nonumber \\
  & f_i \ge 0 \;\; \text{for } i \in \{1...N_c\}
  \label{eq::DualCE}
\end{align}

In \eqref{eq::DualCE}, $w^*_{ei}(f_i)$ is the convex conjugate \cite{Rockafellar2015}
of the elastic strain energy function \eqref{eq::Constitutive2}.
For power-law elastic constitutive laws of the form $w_{ei}(c_i) = \frac{1}{p}k_{ei} c_i^p$,
which encompass Hookean and Hertzian contact laws for some elastic stiffness $k_{ei}$,
the convex conjugate of $w_{ei}(c_i)$ is
\begin{equation}
  w^*_{ei}(f_i) = \left(\frac{p-1}{p}\right)  k_{ei}^{\frac{1}{1-p}}  f_i^{\frac{p}{p-1}}
  \label{eq::ComplementaryEnergy}
\end{equation}
For Hookean contacts, $p=2$, and \eqref{eq::ComplementaryEnergy} reduces
to $w^*_{ei}(f_i) = \frac{1}{2 k_{ei}} f_i^2$.
For Hertzian contacts, $p=\frac{5}{2}$, leading to
$w^*_{ei}(f_i) = \frac{3}{5}k_{ei}^{-\frac{2}{3}} f_i^{\frac{5}{3}}$.

Up to this point, we have formulated a no-tension truss for truss members
of finite stiffness, taking advantage of the equivalence between the
potential energy and complementary energy formulations of the problem.
Now, we consider taking a the limit of infinitely rigid truss members
and restrict ourselves to only power-law constitutive laws with complementary
energy of the form \eqref{eq::ComplementaryEnergy}.

Let the elastic stiffness of each truss member $k_{ei}$ be composed of two
parts, a \textit{relative} stiffness $\tilde{k}_i > 0$ and a scale $\kappa > 0$.
The scale $\kappa$ can be interpreted as a ``representative stiffness''
of a contact.
The relative stiffness thus represents the ratio of a contact stiffness
to the representative stiffness $k_{ei}/\kappa$ such that
\begin{equation}
  k_{ei} = \kappa \tilde{k}_i
  \label{eq::stiffnessDecomposition}
\end{equation}
By performing this decomposition and taking the limit as
$\kappa \rightarrow \infty$, we can take the limit of all particle
stiffnesses approaching infinity while maintaining some desired
stiffness ratios among the contacts.
We expect this approach to work best when $\tilde{k}_i$ is O(1)
for all contacts in a network, which is reasonable when all grains
are made of a similar class of materials, such as a combination of
various metallic grains together.

This decomposition allows us to factor
$\kappa^{\frac{1}{1-p}}$ out of the objective function in \eqref{eq::DualCE}.
Since a multiplicative constant does not affect the location of the
optimal solution to \eqref{eq::DualCE}, it can be dropped entirely, resulting
the following ``scaled'' optimization problem.

\begin{align}
  \min_{f} \;& \sum_{i=1}^{N_c}
  \left(\frac{p-1}{p}\right)  \tilde{k}_i^{\frac{1}{1-p}}  f_i^{\frac{p}{p-1}} \nonumber \\
  s.t. \;& (B^T f)_j = \tilde{F}_j \;\; \text{for } j \in \Gamma_N \nonumber \\
  & f_i \ge 0 \;\; \text{for } i \in \{1...N_c\}
  \label{eq::DualCEScaled}
\end{align}

Now, we can take the limit as $\kappa \rightarrow \infty$, and \eqref{eq::DualCEScaled}
is unaffected.
Thus, an optimal solution to \eqref{eq::DualCEScaled} results in the set of
forces that arise in the limit of infinite stiffness of an elastic truss,
which is precisely the condition for compatible forces assuming
simultaneously-resolved contacts.
Note that the minimization problem (\ref{eq::DualCEScaled}) plays a role 
similar to the role of Eq \eqref{eq::ElasticCompatibility} in compatible elasticity.
The last remaining piece to use this zero-tension truss formulation
is to define the external load $\tilde{F}$, which contains the
impulses that contacts must deliver to particles in order to satisfy
the no-penetration condition of CD.

\subsection{Applying the No-Tension Truss to CD}

To apply this to CD, we have to exploit a feature of frictionless
CD: uniqueness of velocities.
Although the contact forces are not unique, the velocity solution
is unique.
Since the velocities at the end of the timestep are unique,
we can treat them as known.
Thus, the equations of motion used to update velocities for the system
resemble an equilibrium condition for a zero-tension truss subject
to inertial loading.
After rearrangement of \eqref{eq::EoMDiscrete}, we find a set of
equilibrium conditions
\begin{equation}
  B^T f = \frac{1}{\Delta t}M\left(v^\tdt - v^t\right) - F^{ext} = \tilde{F}
  \label{eq::MasonryEquilibrium}
\end{equation}
where $\tilde{F}$ is a generalized external body force that includes the
known contribution from the change of inertia during a time step.
Thus, solving \eqref{eq::DualCEScaled} with this generalized body force
will result in a set of contact forces that solve CD and satisfy the conditions
for being the solution of a zero-tension truss.

\section{Two-Step Solution Algorithm}
Putting together the results from the previous section,
we propose a two-step solution algorithm for finding compatible
contact forces for CD, laid out in pseudo-code
in Algorithm \ref{alg::SolutionAlgorithm}.
First, solve the complementarity problem \eqref{eq::ComplementarityFinal}
either directly or in its QP form \eqref{eq::CDQP} using any appropriate
solver.
In a following section, we give numerical examples of solving 
\eqref{eq::ComplementarityFinal} directly using a Projected
Gauss-Seidel (PGS) solver \cite{Morales2008} and of solving
\eqref{eq::CDQP} using an Accelerated Projected Gradient Descent (APGD)
solver \cite{Mazhar2015}.
From this indeterminate solution, compute the updated velocities, and
use these to subsequently compute the generalized body force as shown
in \eqref{eq::MasonryEquilibrium}.
Finally, solve the no-tension truss QP/NLP \eqref{eq::DualCEScaled}, the
solution of which is a set of compatible contact forces that represent
the solution of an elastic contact network in the limit of infinite stiffness.
To solve \eqref{eq::DualCEScaled}, we implemented an Augmented Lagrangian
solver based on the description of the LANCELOT solver \cite{Wright1999} \cite{Conn2013}.
We re-use our APGD solver from the CD step to solve the subproblems
arising from the Augmented Lagrangian algorithm, as it is well-suited
to solve arbitrary bound-constrained convex optimization problems.

Since the compatibility correction is entirely separate from the 
CD LCP solve, this method can be integrated into existing codes with minimal
effort.

\begin{algorithm}[H]
  \caption{
    Time step update using the
    two-step compatible CD algorithm
  }
  \label{alg::SolutionAlgorithm}
  \begin{algorithmic}
    \State $\text{\textbf{Step 1: Solve Classic CD}}$
    \State $\text{Compute contacts, gap function, gap gradient:}$
    \State $\Phi(q^t) = CollisionDetection(q^t,v^t)$ 
    \State $B = \frac{\partial \Phi}{\partial q^t}$
    \State
    \State $\text{Set up and solve the CD QP \eqref{eq::CDQP}:}$
    \State $M^{-1} = UpdateInverseMassMatrix(q^t)$
    \State $N = \Delta t BM^{-1}B^T$
    \State $p = \frac{\Phi}{\Delta t} + Bv^t + \Delta tBM^{-1}F^{ext}$
    \State $f = SolveCDQP(N,p)$
    \State
    \State $\text{Update Velocity, Position:}$
    \State $v^\tdt = v^t+\Delta t M^{-1}\left(B^T f + F^{ext}\right)$
    \State $q^\tdt = q^t + \Delta t L(q^t)v^\tdt$
    \State
    \State $\text{\textbf{Step 2: Compatible CD Post-processing}}$
    \State $\text{Compute the generalized body force required for equilibrium \eqref{eq::MasonryEquilibrium}:}$
    \State $\tilde{F} = \frac{1}{\Delta t}M(v^\tdt - v^t) - F^{ext}$
    \State
    \State $\text{Solve the No-tension truss QP/NLP
      \eqref{eq::DualCEScaled}}:$
    \State $f^{Compatible} = SolveCompatibilityQP/NLP(\{\tilde{k}_i\}, B,\tilde{F})$
  \end{algorithmic}
\end{algorithm}

\section{Numerical Experiments}
This section contains examples of indeterminate configurations that
demonstrate the efficacy of the proposed two-step algorithm for finding compatible
contact forces.
We compare the output of this algorithm to the contact forces generated
by solving \eqref{eq::ComplementarityFinal}, which we refer to as ``classic CD.''
In each of these tests, the same configuration is simulated using
classic CD solved using multiple algorithms, compatible CD, and a
penalty-based DEM.
The geometries are chosen to provide an effective demonstration of
the compatible CD algorithm's ability to reproduce the desired contact
force behavior where classic CD algorithms fail to do so.

We considered two algorithms for solving the classic CD problem.
First, we implemented a Projected Gauss-Seidel (PGS) solver 
that operates directly on the complementarity form of the problem,
\eqref{eq::ComplementarityFinal}.
The PGS solver has been well-studied and is a popular method for solving
linear complementarity problems \cite{Morales2008}, \cite{Cottle2009}, \cite{Radjai2009}.
We also implemented an Accelerated Projected Gradient Descent (APGD)
solver \cite{Mazhar2015}, which is a gradient-based method based on
Nesterov's accelerated gradient descent \cite{Nesterov1983}.
The APGD solver operates on \eqref{eq::CDQP}, the QP form of CD.


\subsection{Indeterminately-supported Beam}
This two-dimensional configuration consists of a single rigid platform
of length $L$ supported in the $y$ direction by three fixed spherical 
supports located at $x=0$, $x=L/2$, and $x=L$.
The platform is loaded by a body force acting in the $-y$ direction,
resulting in a total weight of $w=10$.
It is dropped from a height of 1\% of a particle radius,
and all images are generated when the system has settled to a static configuration.

Due to the indeterminacy of the problem in the rigid limit, solution algorithms
are not guaranteed to find the correct static forces unless they consider
a compatibility condition,
even in the case where all supports have the same relative stiffness.
Indeed, we found that the PGS solver produced results that depend
strongly on the order in which contacts were visited algorithmically.

\subsubsection{Uniform Contact Stiffness}
In the first set of simulations, all contacts have the same relative stiffness,
and the stiffness scale was allowed to go to infinity.
In this case, the analytical solution is for all three contacts to carry
the same force, which is $f = 10/3$ in this case.

\begin{figure}[H]
  \centering
  \begin{subfigure}{.45\textwidth}
    \includegraphics[width=\textwidth]{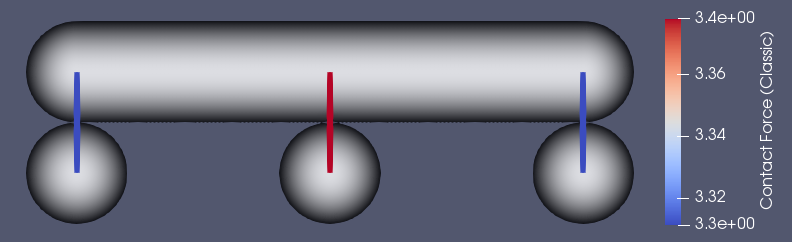}
    \caption{}
    \label{fig::SupportedBeamPGS}
  \end{subfigure}
  \begin{subfigure}{.45\textwidth}
    \includegraphics[width=\textwidth]{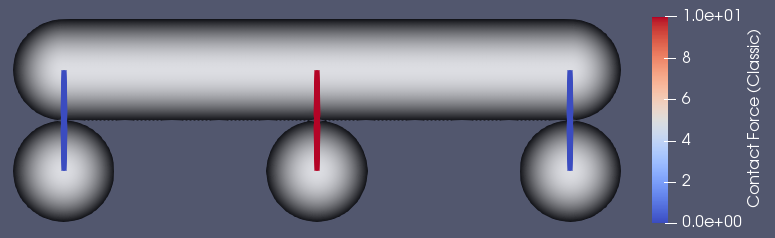}
    \caption{}
    \label{fig::SupportedBeamPGSReordered}
  \end{subfigure}
  \caption{
    (a)
    The PGS algorithm results in a non-equal distribution of
    forces. The distribution is sensitive to the numbering of
    the contacts.
    (b)
    The contacts were reoredered so that the middle contact
    was the first one visited in the PGS solver. The result
    is an unrealistic force distribution where the middle contact
    carries 100\% of the platform weight and the outer contacts
    carry nothing.
    This is a pathological example of how indeterminacy can
    affect the contact force distribution.
  }
\end{figure}

\begin{figure}[H]
  \centering
  \includegraphics[width=.5\textwidth]{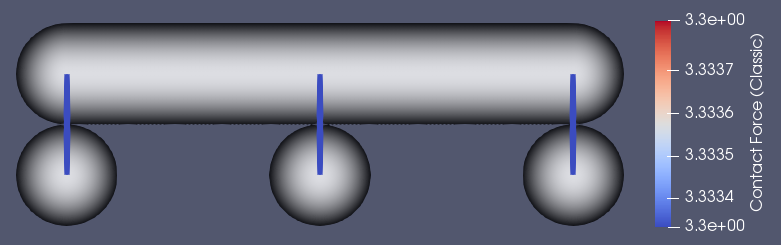}
  \caption{
    Solving classic CD with the APGD solver results
    in the correct solution in this instance, dividing the
    platform weight evenly across the three contacts.
  }
  \label{fig::SupportedBeamAPGD}
\end{figure}

\begin{figure}[H]
  \centering
  \includegraphics[width=.5\textwidth]{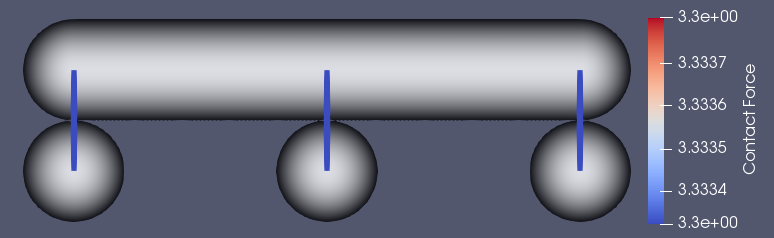}
  \caption{
    Solving compatible CD with results
    in the correct solution, dividing the
    platform weight evenly across the three contacts.
  }
  \label{fig::SupportedBeamMasonry}
\end{figure}

\subsubsection{Non-Uniform Contact Stiffness}
In this next set of simulations, the contact between the middle particle and the platform
is has a relative stiffness $\tilde{k} = 3$, and the outer contacts
have a relative stiffness $\tilde{k} = 1$.
The analytical solution in this setup is for the middle contact
to carry three times the weight of either outer contact.
Since the platform has a weight of 10, the middle contact should
have $f = 6$ and the outer contacts should each have $f = 2$.
Since classic CD has no notion of contact stiffnesses, only the
compatible CD results are shown here.
The classic CD results are identical to the cases shown above,
so classic CD is (as expected) unable to predict the effects
of contact stiffness variation.

\begin{figure}[H]
  \centering
  \includegraphics[width=.5\textwidth]{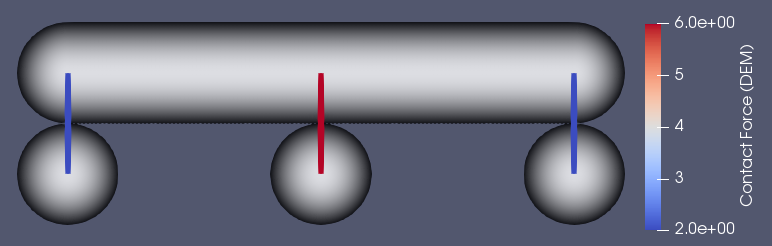}
  \caption{
    The reference solution computed by DEM shows a non-uniform
    distribution of forces across the contacts due to the
    non-uniform contact stiffness. Recall that the middle
    contact is three stimes as stiff as the outer contacts.
  }
  \label{fig::SupportedBeamMultistiffnessDEM}
\end{figure}

\begin{figure}[H]
  \centering
  \includegraphics[width=.5\textwidth]{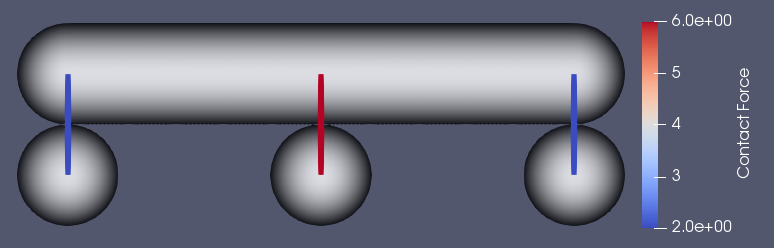}
  \caption{
    For the multi-stiffness case, compatible CD
    correctly predicts the non-uniform force distribution
    across the three contacts. Recall that the middle
    contact is three times as stiff as the outer contacts.
    This solution matches the reference solution in
    figure \ref{fig::SupportedBeamMultistiffnessDEM}.
  }
  \label{fig::SupportedBeamMultistiffnessMasonry}
\end{figure}

\subsection{Cannonball Packing}

\begin{figure}[H]
  \centering
  \includegraphics[width=0.4\textwidth]{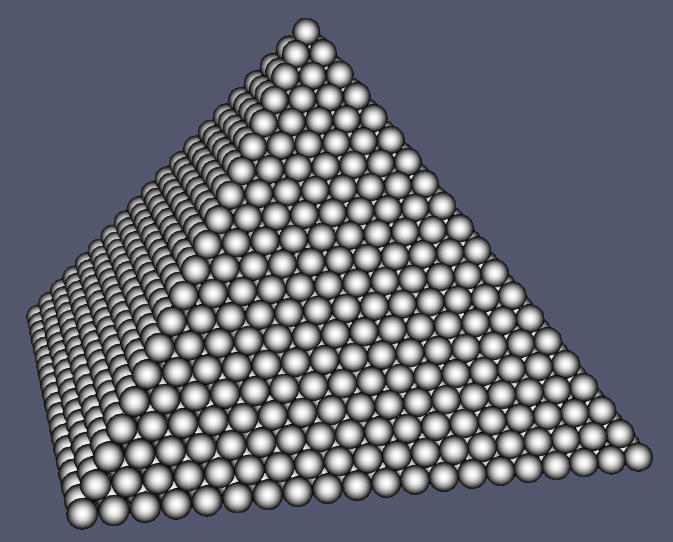}
  \caption{
    Static configuration of the sphere pyramid.
  }
  \label{fig::CannonballConfig}
\end{figure}

The first example problem is a square-based pyramid of close-packed spheres
with 20 spheres along the base, shown in figure \ref{fig::CannonballConfig}.
In this configuration, the interior particles have twelve neighbors, resulting
in indeterminacy in the contact forces.
The particles are initially separated vertically from their neighbors
by 5\% of a radius and dropped simultaneously under a gravitational body force,
holding the base layer fixed to form the pyramid.
In the present set of simulations, the spheres are $2.5mm$ in diameter with density $2000kg/m^3$.
For the DEM comparisons that follow, the spheres were given a Young's modulus of
50 GPa, an approximation of glass beads,
and a viscous damping $\gamma = 2\sqrt{km}$ corresponding to restitution
coefficient $e_n = 0$.
The viscoelastic Hookean and Hertzian constitutive laws for DEM contact forces
are given in \cite{Silbert2001}, with contact stiffness $k$ computed
from Young's modulus as in \cite{Zhang2005}.
The CD simulations were run using a conservatively small time step of 0.001 seconds.
The DEM simulations were run using a time step of 5.0e-7 seconds, which was close
to the stable time step.

\begin{figure}[H]
  \centering
  \begin{subfigure}{.35\textwidth}
    \includegraphics[width=\textwidth,trim=.0in 1.5in 0in 1in, clip]{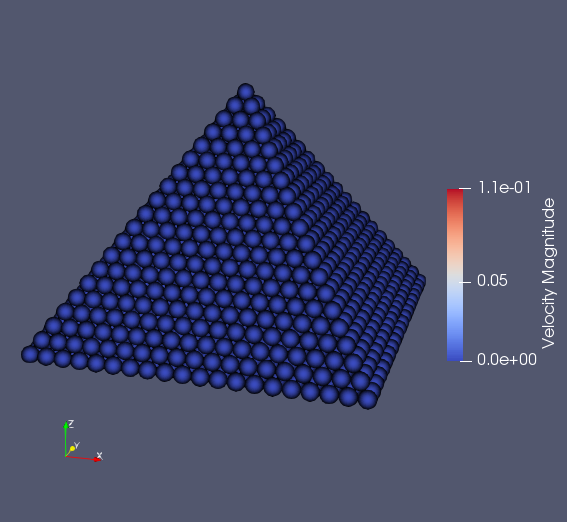}
    \caption{}
  \end{subfigure}
  \begin{subfigure}{.35\textwidth}
    \includegraphics[width=\textwidth,trim=.0in 1.5in 0in 1in, clip]{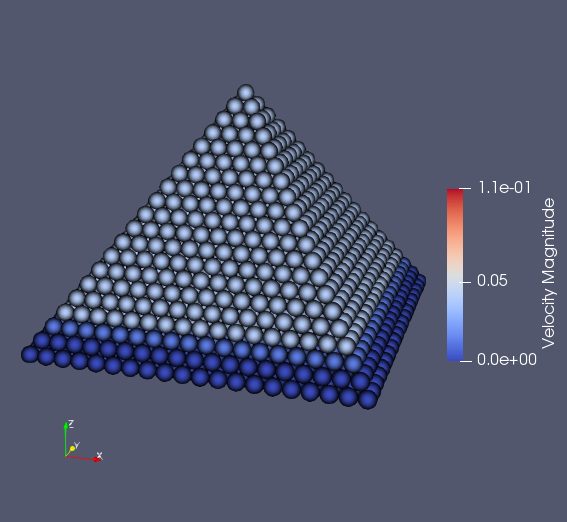}
    \caption{}
  \end{subfigure}
  \begin{subfigure}{.35\textwidth}
    \includegraphics[width=\textwidth,trim=.0in 1.5in 0in 1in, clip]{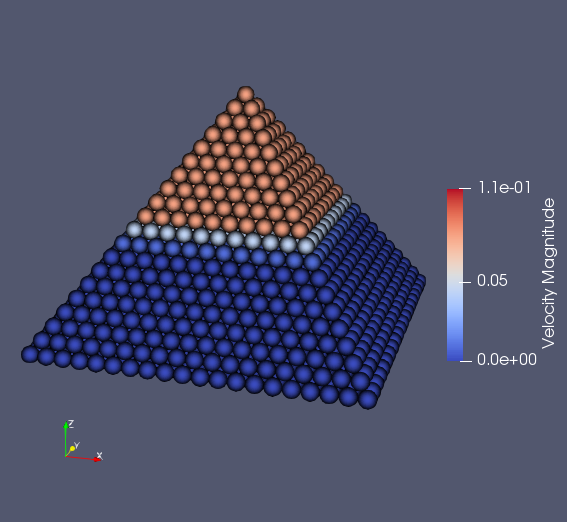}
    \caption{}
  \end{subfigure}
  \begin{subfigure}{.35\textwidth}
    \includegraphics[width=\textwidth,trim=.0in 1.5in 0in 1in, clip]{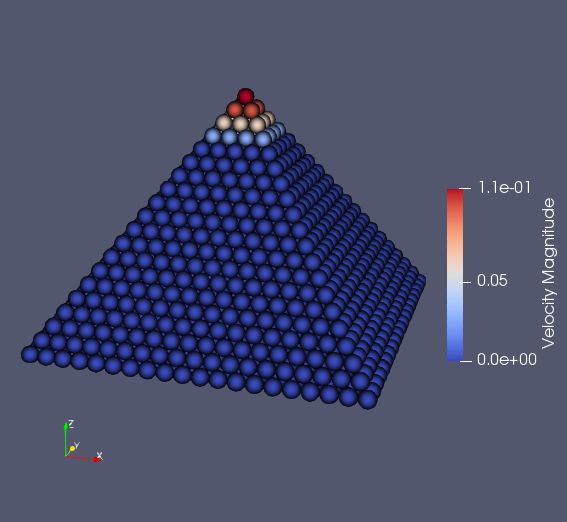}
    \caption{}
  \end{subfigure}
  \caption{Particle velocity magnitudes at four different times of 
    CD simulation during settling.
    (a) $t = 0$,
    (b) $t = 0.004$,
    (c) $t = 0.008$,
    (d) $t = 0.016$
  }
  \label{fig::CannonballSequence}
\end{figure}

\subsubsection{Cannonball Packing: Uniform Stiffness Hookean Contacts}
\begin{figure}[H]
  \centering
  \begin{subfigure}{.4\textwidth}
    \includegraphics[width=\textwidth]{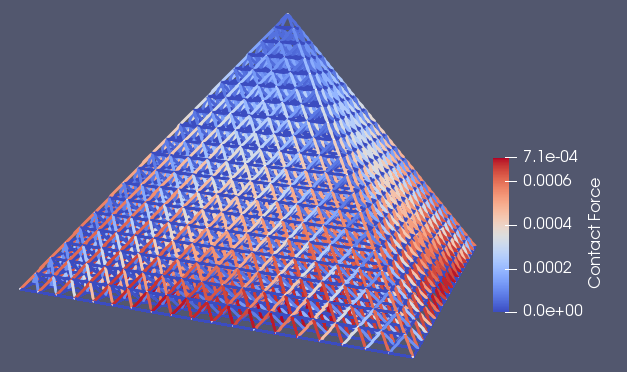}
    \caption{}
    \label{fig::CannonballZeroRestitutionDEM}
  \end{subfigure}
  \begin{subfigure}{.4\textwidth}
    \includegraphics[width=\textwidth]{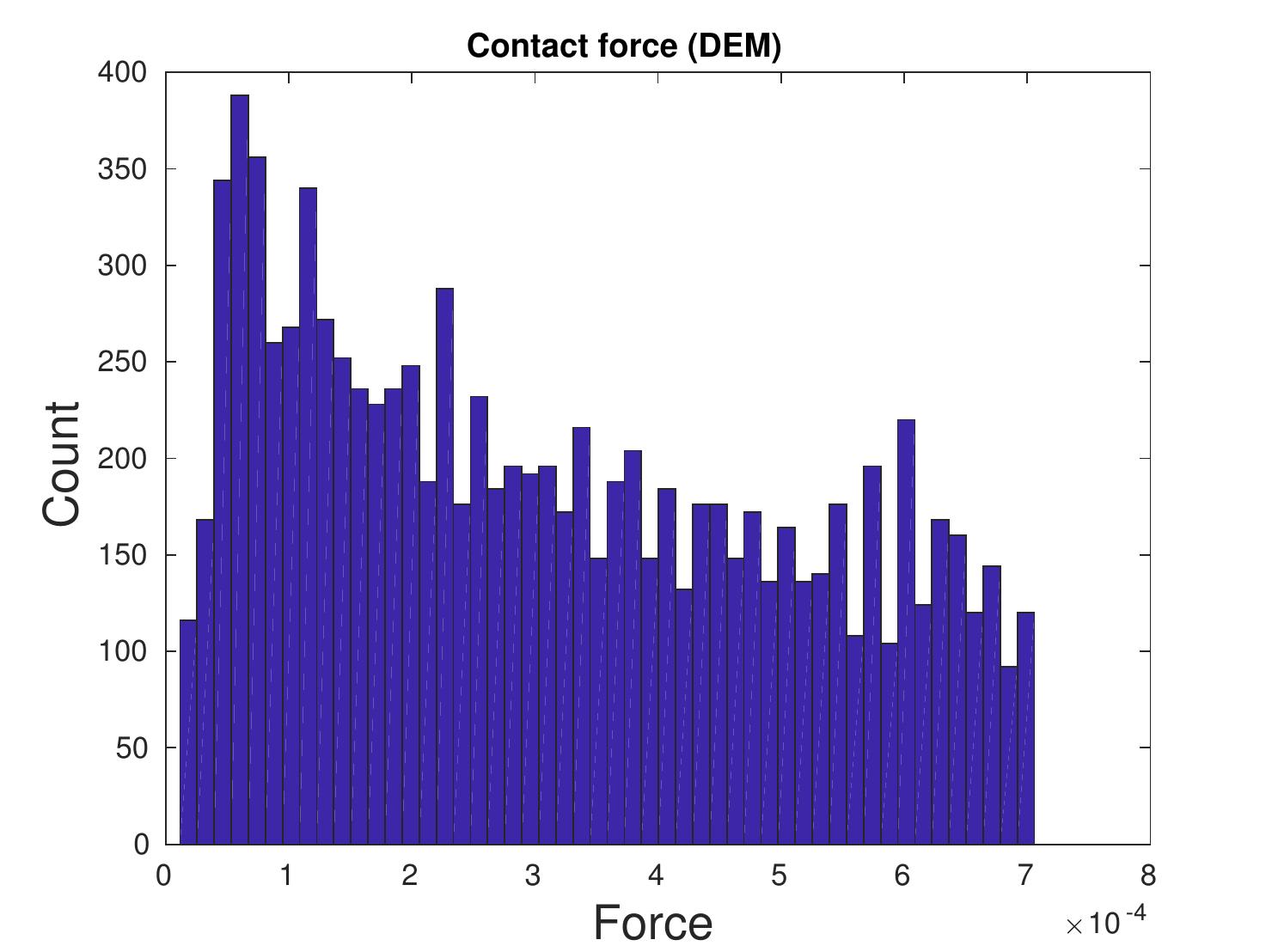}
    \caption{}
    \label{fig::CannonballZeroRestitutionDEMHist}
  \end{subfigure}
  \caption{
    (a)
    The reference solution for the contact forces is given by DEM with Hookean
    contacts.
    (b) Histogram of the nonzero contact forces in the cannonball packing.
  }
\end{figure}

\begin{figure}[H]
  \centering
  \begin{subfigure}{.4\textwidth}
    \includegraphics[width=\textwidth]{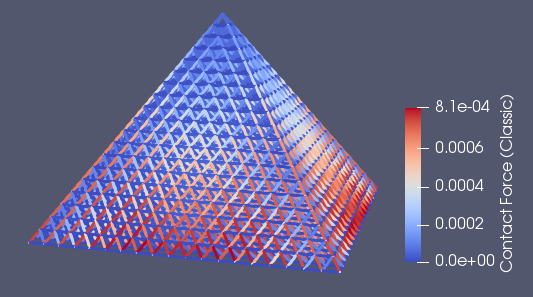}
    \caption {}
    \label{fig::CannonballZeroRestitutionCDAPGD}
  \end{subfigure}
  \begin{subfigure}{.4\textwidth}
    \includegraphics[width=\textwidth]{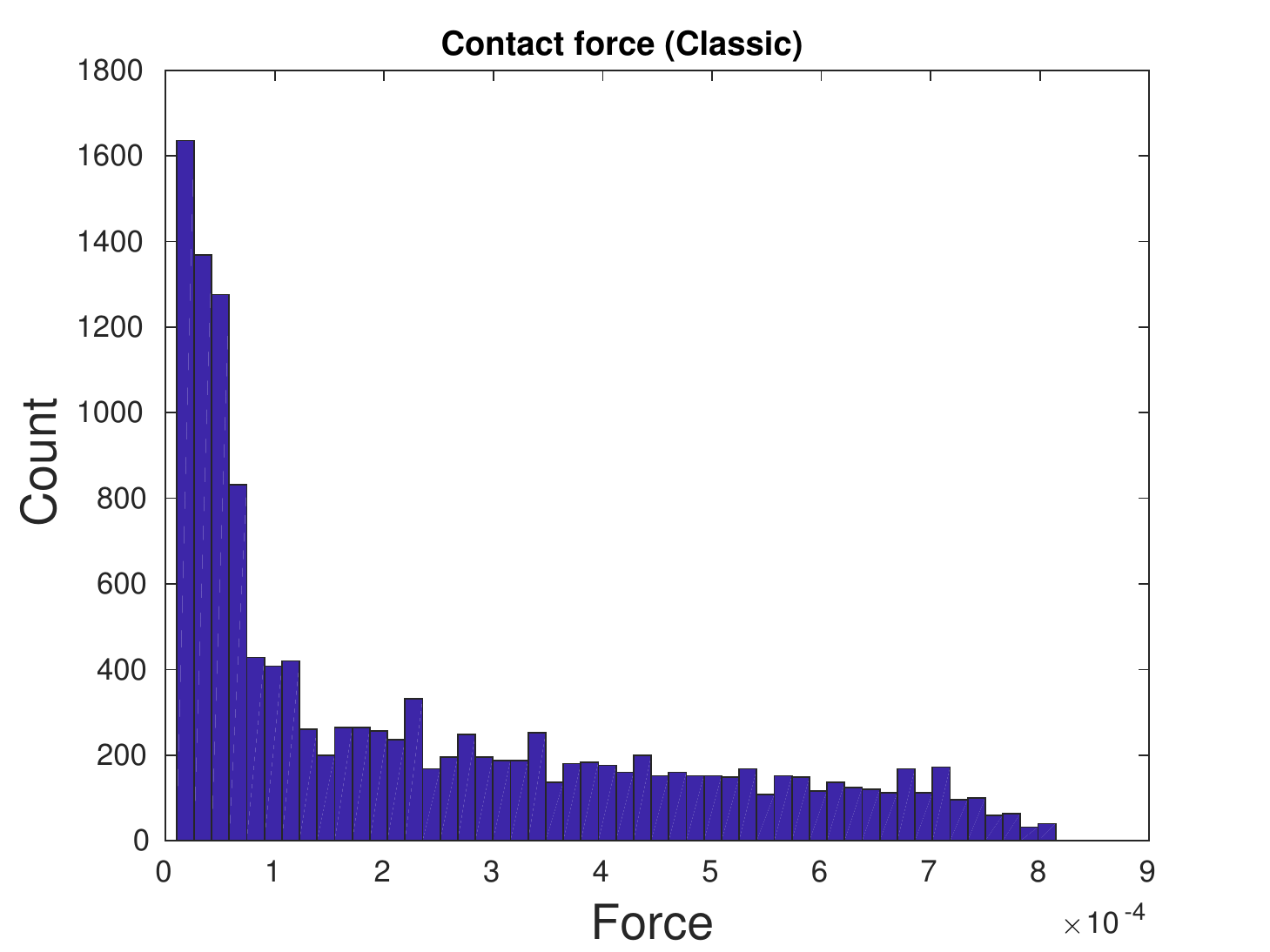}
    \caption{}
    \label{fig::CannonballZeroRestitutionCDAPGDHist}
  \end{subfigure}
  \caption{
    (a)
    The force distribution of classic CD using the APGD solver is qualitatively similar
    to the reference DEM solution in figure \ref{fig::CannonballZeroRestitutionDEM},
    but the maximum contact forces are over-predicted by approximately 14\%.
    (b)
    The histogram of nonzero contact forces shows more clearly the significant disagreement
    from the reference solution, especially in the low range, where CD has found
    a solution with many more load-bearing contacts.
  }
\end{figure}

Once the dropped pyramidal pile has become static, we find that classic CD is unable to reliably replicate the reference DEM solution,
shown in figure \ref{fig::CannonballZeroRestitutionDEM}, even though it is in the
set of feasible solutions.
The APGD solver produces qualitatively correct results, but is unable to quantitatively
match the reference force distribution.
The force distribution found by the APGD solver is shown in figure
\ref{fig::CannonballZeroRestitutionCDAPGD}.
The PGS solver produces qualitatively incorrect force distributions, breaking the
symmetry of the problem due to the sensitivity on contact ordering.
This is a well-known effect of PGS solvers, and has been studied in more detail in
the context of simultaneous impact problems in \cite{Smith2012}.

\begin{figure}[H]
  \centering
  \begin{subfigure}{.4\textwidth}
    \includegraphics[width=\textwidth]{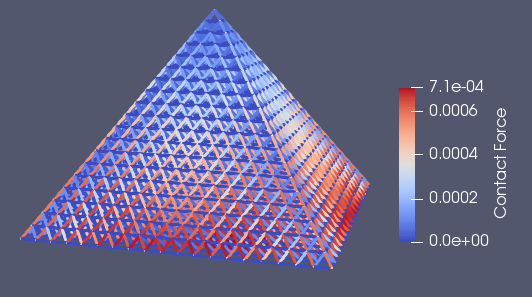}
    \caption{}
    \label{fig::CannonballZeroRestitutionCDMasonry}
  \end{subfigure}
  \begin{subfigure}{.4\textwidth}
    \includegraphics[width=\textwidth]{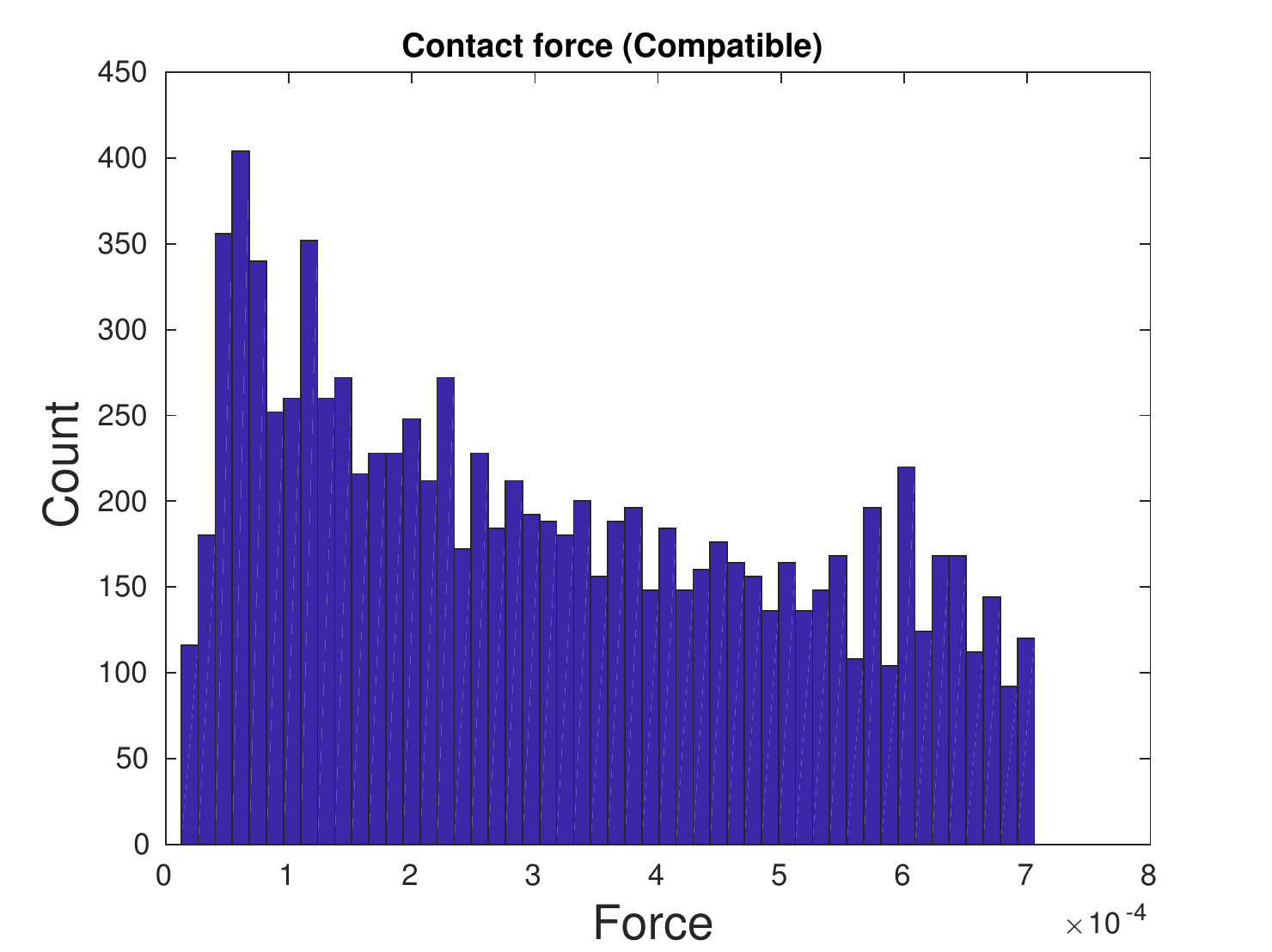}
    \caption{}
    \label{fig::CannonballZeroRestitutionCDMasonryHist}
  \end{subfigure}
  \caption{
    (a) The force distribution of the compatible CD algorithm matches the reference force
    distribution computed by DEM in figure \ref{fig::CannonballZeroRestitutionDEM},
    regardless of the algorithm used to solve the initial CD LCP.
    (b) The histogram confirms the near-perfect match between the contact forces
    shown here and those in the reference problem. The quality of the match
    is determined by the tolerances to which the CD and compatiblility QPs are solved.
  }
\end{figure}

Regardless of the solution algorithm used to solve the classic CD problem,
our post-processing step that finds the compatible solution is consistently
able to replicate the reference DEM force distribution.
Figure \ref{fig::CannonballZeroRestitutionCDMasonry} shows the compatible
contact forces computed by post-processing the CD solution from the APGD algorithm.
The post-processed solution from the PGS algorithm is identical, so it has
been omitted.

\begin{figure}[H]
  \centering
  \begin{subfigure}{.4\textwidth}
    \includegraphics[width=\textwidth]{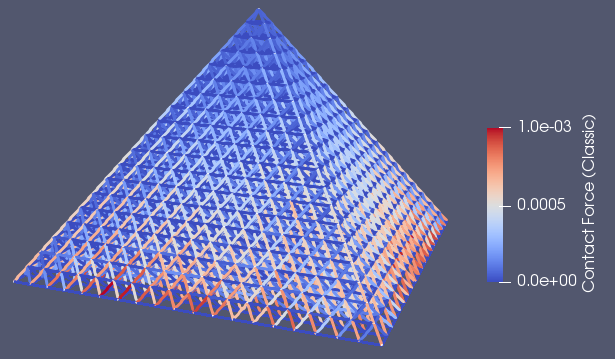}
    \caption{}
    \label{fig::CannonballZeroRestitutionCDPGS}
  \end{subfigure}
  \begin{subfigure}{.4\textwidth}
    \includegraphics[width=\textwidth]{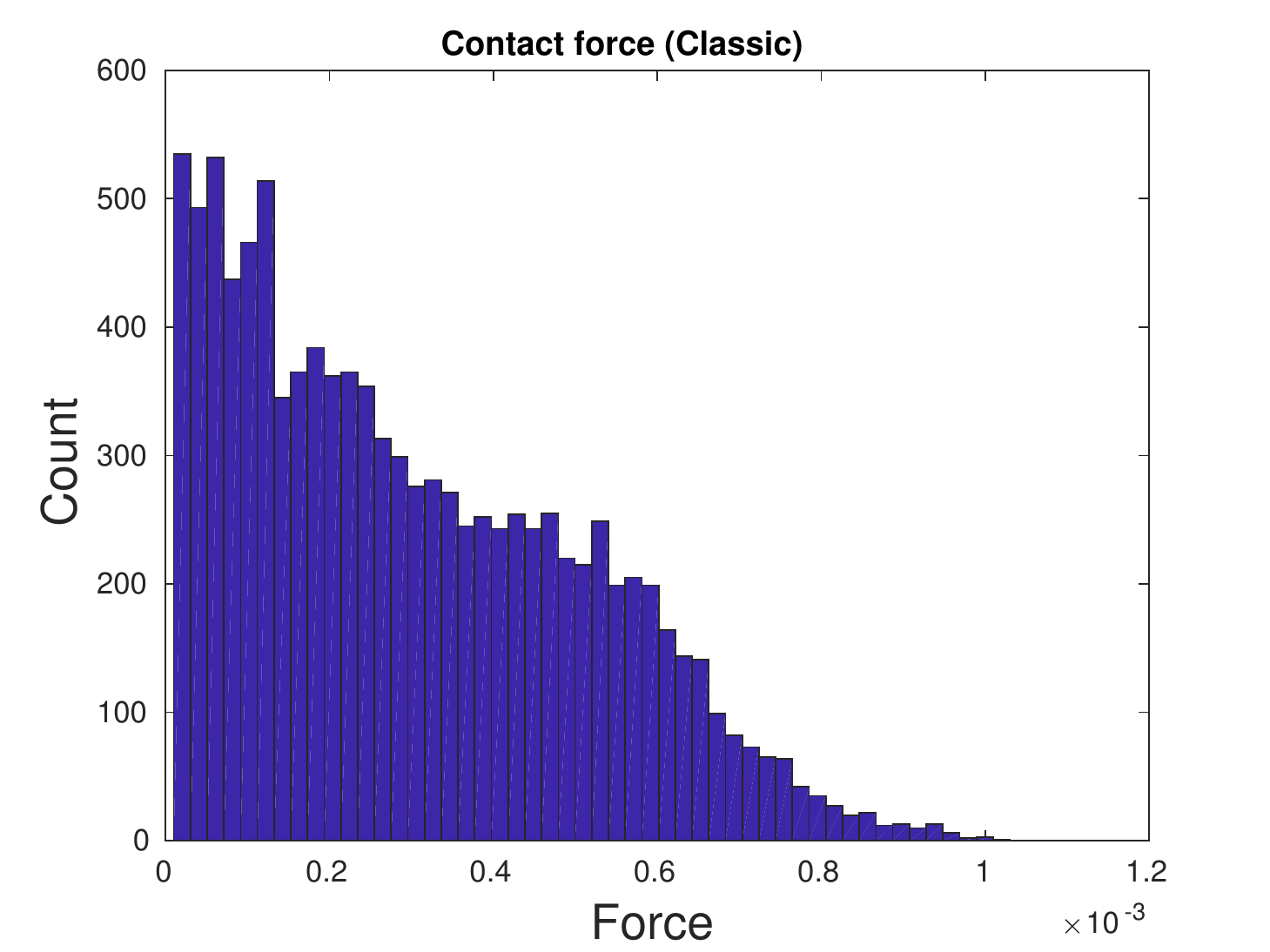}
    \caption{}
    \label{fig::CannonballZeroRestitutionCDPGSHist}
  \end{subfigure}
  \caption{
    (a)
    The force distribution of classic CD solved using the PGS algorithm
    fails to preserve the symmetry of the reference solution due to the
    inherent sensitivity to contact ordering in the Gauss-Seidel iteration.
    As is often the case, PGS drastically over-predicts the maximum
    contact force in a system, in this case by approximately 41\%.
    (b)
    The histogram of contact forces confirms the qualitative differences
    in the force distribution from the reference solution in figure
    \ref{fig::CannonballZeroRestitutionDEMHist}.
  }
  
\end{figure}

\begin{figure}[H]
  \centering
  \includegraphics[width=.5\textwidth]{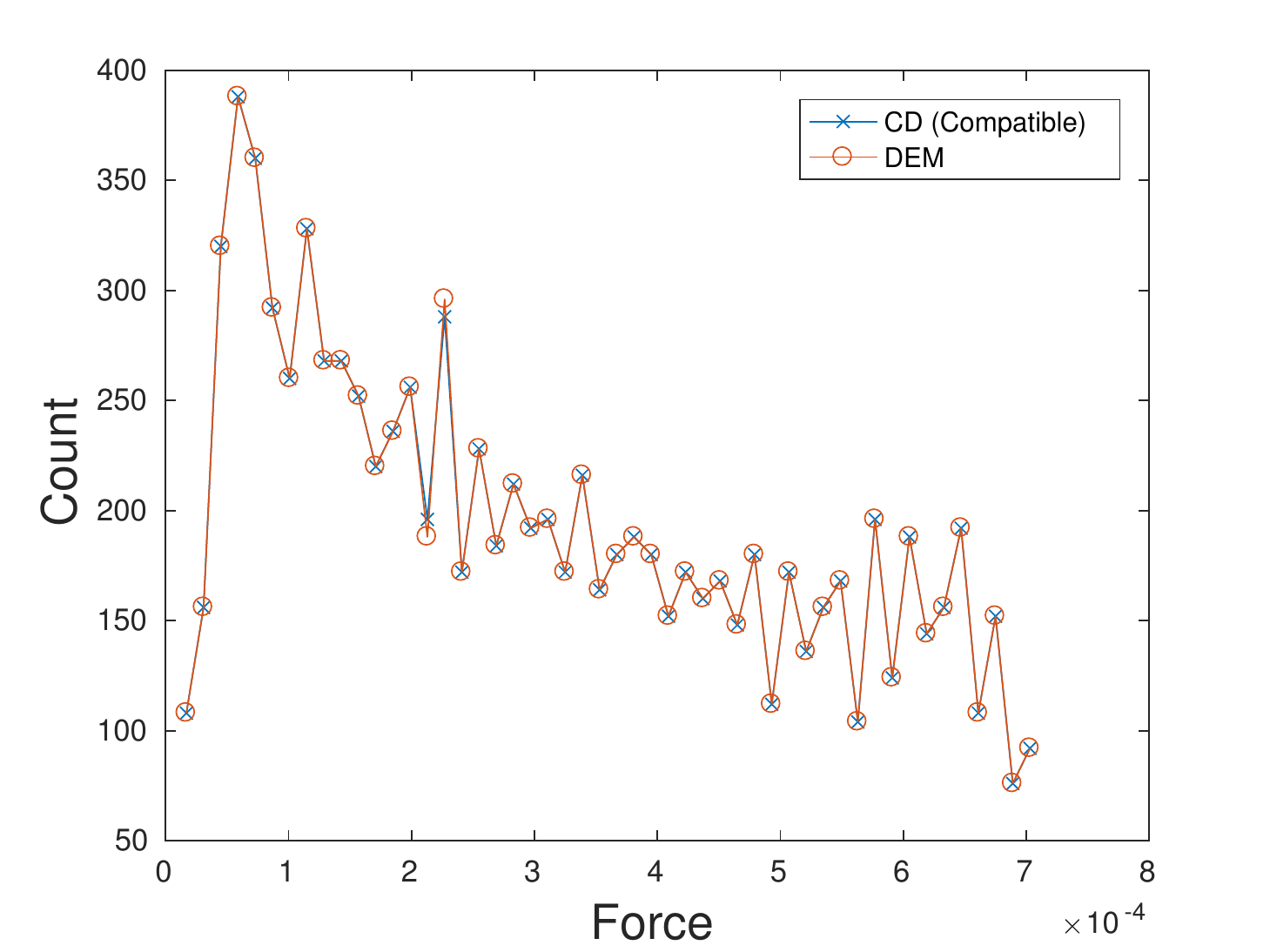}
  \caption{
    Contact force histograms from the DEM reference in figure \ref{fig::CannonballZeroRestitutionDEMHist}
    and compatible CD in figure \ref{fig::CannonballZeroRestitutionCDMasonryHist}
    plotted side-by-side to show more clearly the close agreement between the two
    distributions. The small disagreement in some of the buckets is due to the
    finite solver tolerance in the CD and compatibility problems.
  }
  \label{fig::CombinedCannonballZeroRestitutionAPGDHist}
\end{figure}

\subsubsection{Cannonball Packing: Multi-Species Hookean Contacts}
In the next set of simulations, the material of half of the particles was changed
such that every particle with an even index in the global array had Young's modulus
$E_0$, and odd-indexed particles had Young's modulus $E_1$, where $E_1/E_0 = 5$.
The two particle species result in three possible types of contacts: even-even, even-odd,
and odd-odd, whose relative contact stiffesses were $\frac{1}{5}$, $\frac{1}{3}$,
$1$ respectively.
This is the same scheme used to generate the indeterminately-supported
beam with heterogeneous contact forces described above.

The reason behind this set of simulations was to show that, even in a system
more complicated than the supported beam, force will localize at the stiff
contacts.
As before, this phenomenon is not representable within the standard CD framework,
so we present results only of compatible CD compared against the DEM reference
solution.

\begin{figure}[H]
  \centering
  \begin{subfigure}{.4\textwidth}
    \includegraphics[width=\textwidth]{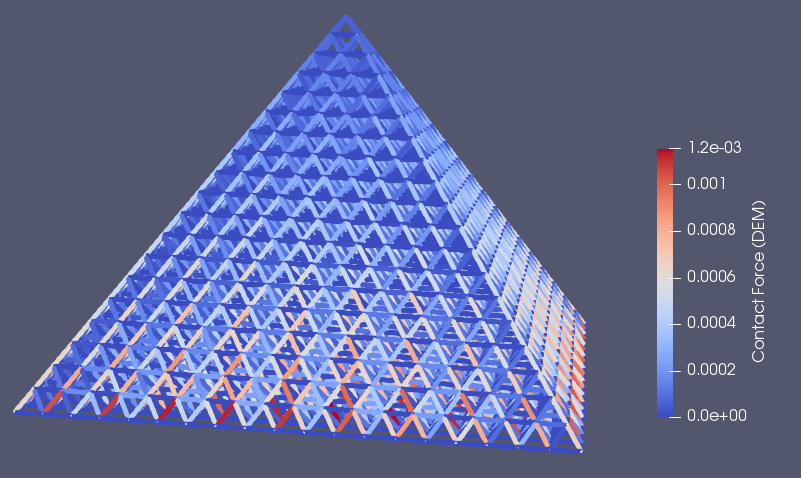}
    \caption{}
    \label{fig::CannonballZeroRestitutionDEMMultiStiffness}
  \end{subfigure}
  \begin{subfigure}{.4\textwidth}
    \includegraphics[width=\textwidth]{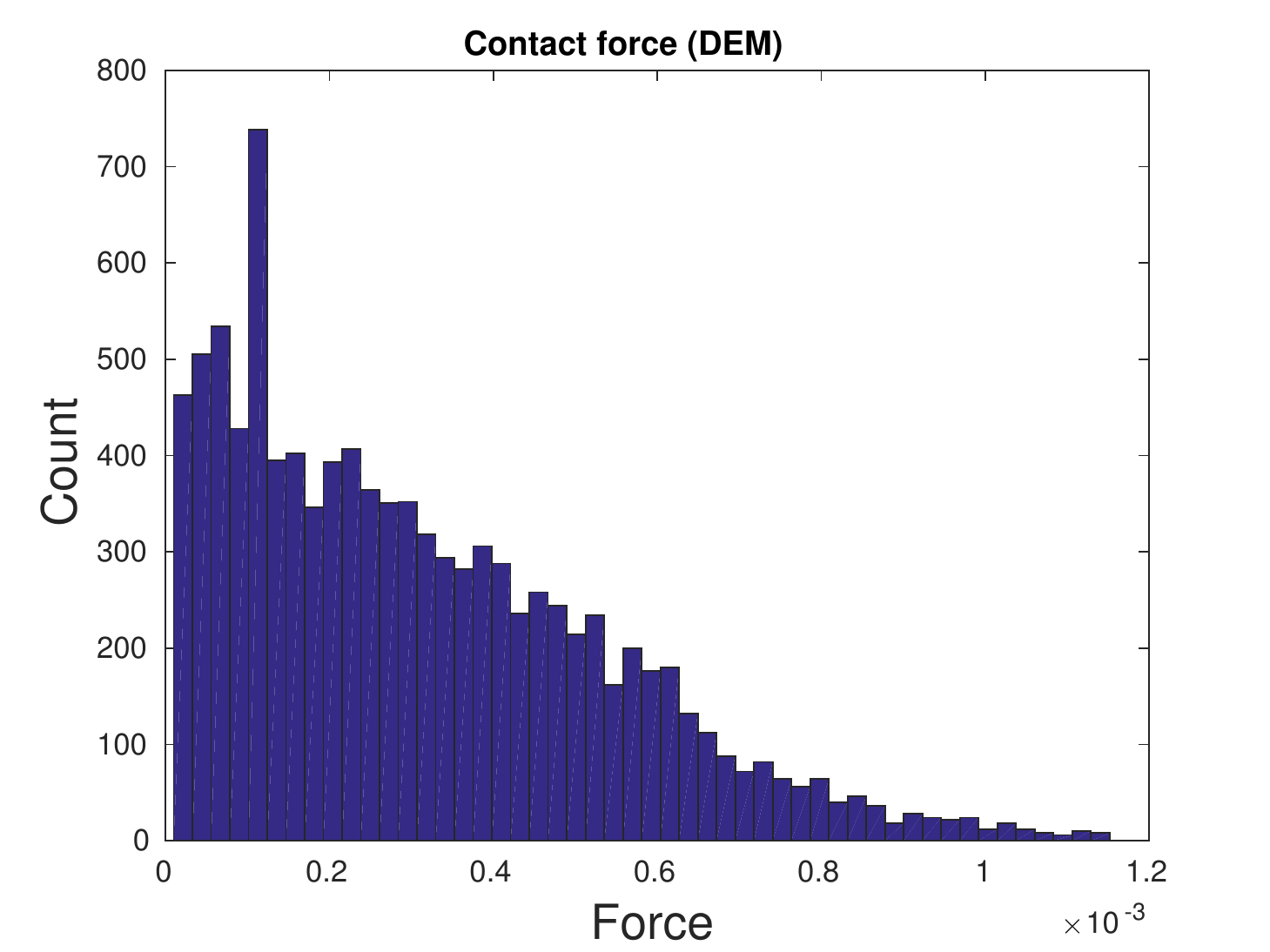}
    \caption{}
    \label{fig::CannonballZeroRestitutionDEMMultiStiffnessHist}
  \end{subfigure}
  \caption{
    (a)
    Reference Hookean DEM solution for the contact forces arising from the 
    cannonball packing with heterogeneous materials.
    The symmetry of the system is broken, and large forces
    localize at the stiff contacts.
    (b) Histogram of the nonzero contact forces in the cannonball packing.
  }
\end{figure}

\begin{figure}[H]
  \centering
  \begin{subfigure}{.4\textwidth}
    \includegraphics[width=\textwidth]{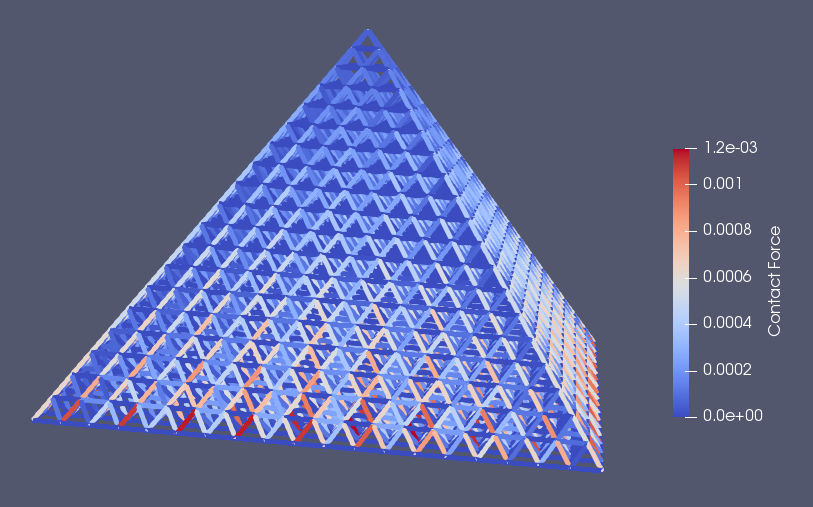}
    \caption{}
    \label{fig::CannonballZeroRestitutionCDMultiStiffnessAPGD}
  \end{subfigure}
  \begin{subfigure}{.4\textwidth}
    \includegraphics[width=\textwidth]{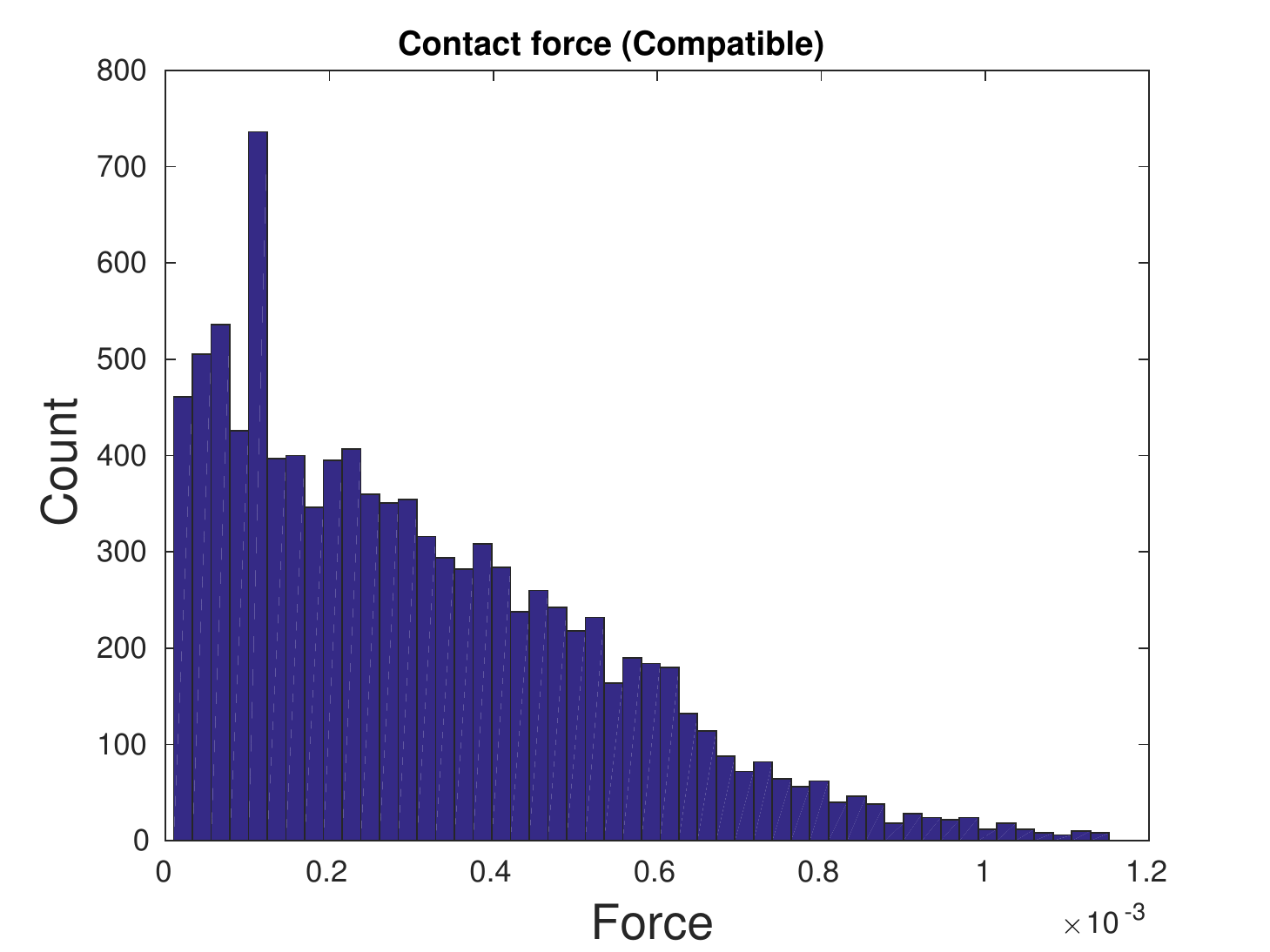}
    \caption{}
    \label{fig::CannonballZeroRestitutionCDMultiStiffnessAPGDHist}
  \end{subfigure}
  \caption{
    (a)
    Contact forces arising from the cannonball packing with heterogeneous
    materials, solved using the compatible CD method for Hookean contacts.
    (b) The histogram confirms the near-perfect match between the DEM and
    compatible CD solutions for this asymmetric, non-trivial force distribution.
  }
\end{figure}

\begin{figure}[H]
  \centering
  \includegraphics[width=.5\textwidth]{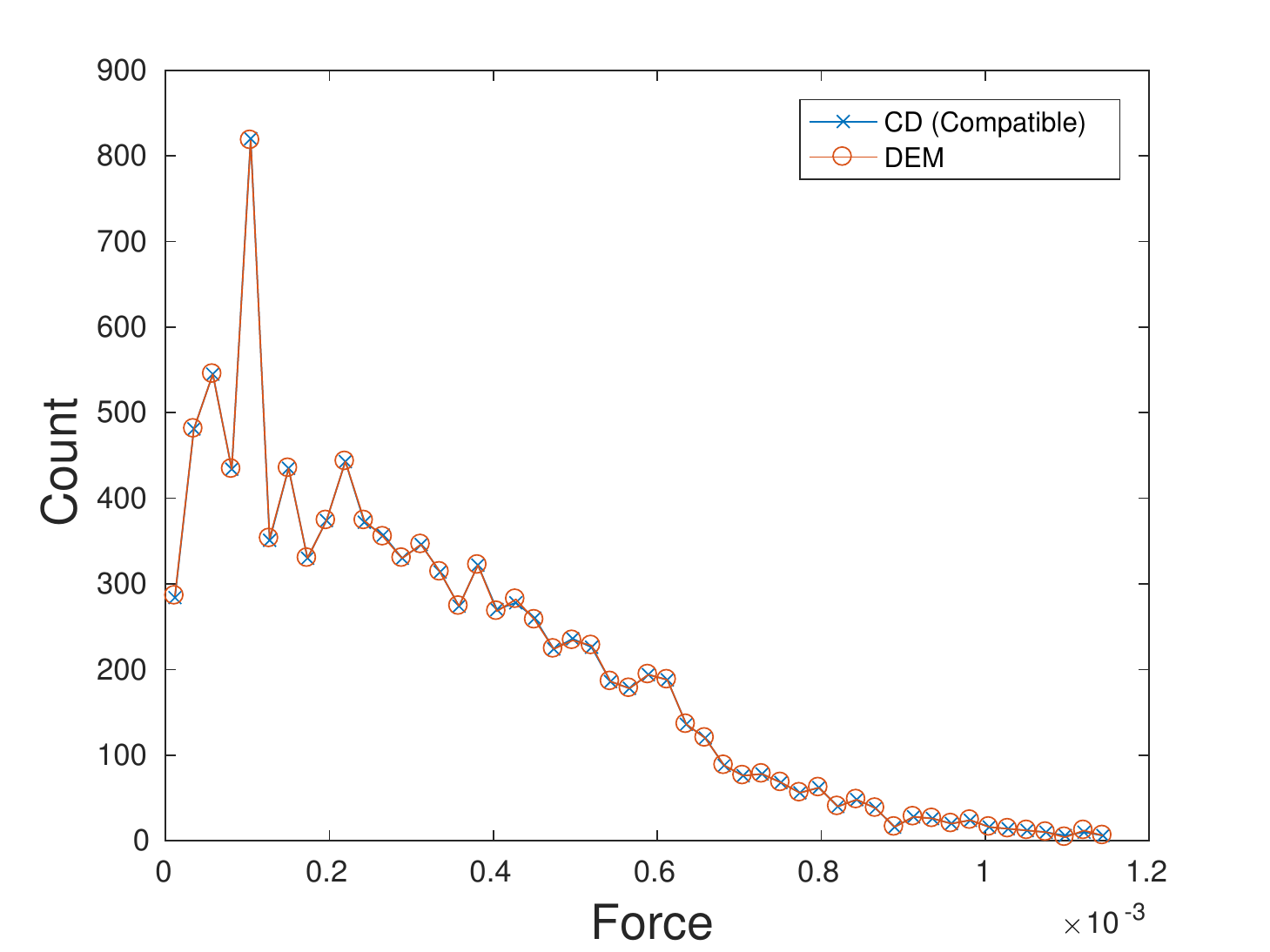}
  \caption{
    Contact force histograms from the DEM reference in figure \ref{fig::CannonballZeroRestitutionDEMMultiStiffnessHist}
    and compatible multi-stiffness CD in figure \ref{fig::CannonballZeroRestitutionCDMultiStiffnessAPGDHist}
    plotted side-by-side to show more clearly the close agreement between the two
    distributions. The small disagreement in some of the buckets is due to the
    finite solver tolerance in the CD and compatibility problems.
  }
  \label{fig::CombinedCannonballZeroRestitutionMultiStiffnessAPGDHist}
\end{figure}

\subsubsection{Cannonball Packing: Uniform Stiffness Hertzian Contacts}
The cannonball simulations were repeated using the
Hertzian contact postprocessing algorithm, which solves \eqref{eq::DualCEScaled} 
corresponding to Hertzian contacts ($p = \frac{5}{2}$) using an augmented Lagrangian algorithm.
Only the DEM and compatible CD results are shown here, since classic CD has no
notion of Hookean vs Hertzian contacts, so the results for classic CD are identical
to those in the previous section.

\begin{figure}[H]
  \centering
  \begin{subfigure}{.4\textwidth}
    \includegraphics[width=\textwidth]{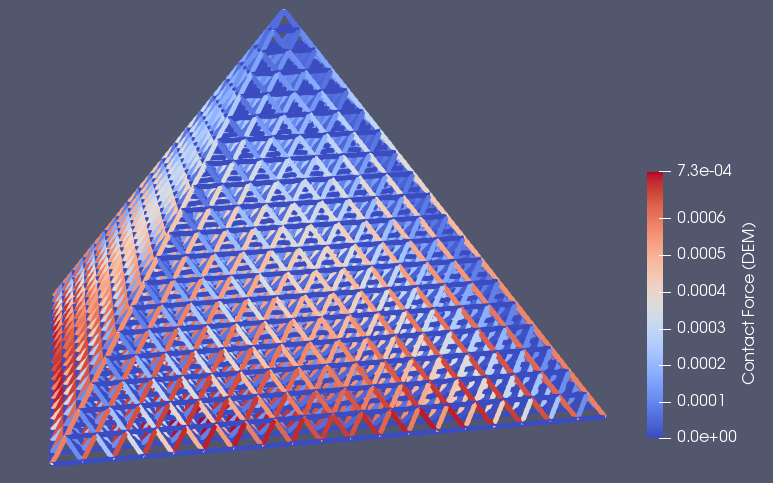}
    \caption{}
    \label{fig::CannonballZeroRestitutionHertzianDEM}
  \end{subfigure}
  \begin{subfigure}{.4\textwidth}
    \includegraphics[width=\textwidth]{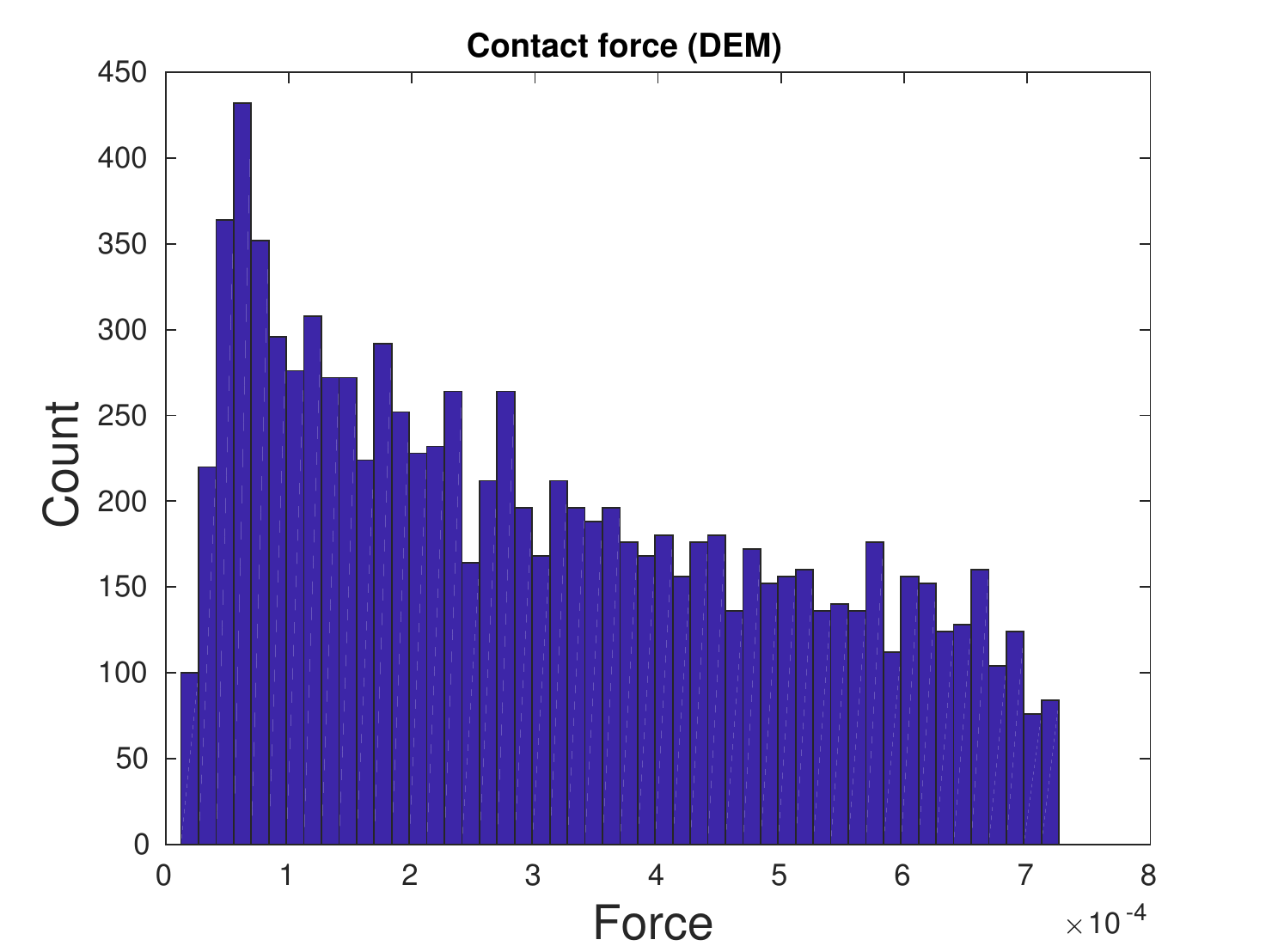}
    \caption{}
    \label{fig::CannonballZeroRestitutionHertzianDEMHist}
  \end{subfigure}
  \caption{
    (a)
    The reference solution for the contact forces is given by DEM with Hertzian
    contacts. Qualitatively, the distribution of contact forces is similar to
    the Hookean case in figure \ref{fig::CannonballZeroRestitutionDEM}, but
    the packing weight is distributed slightly differently, as evidenced
    by the differing maximum forces.
    (b) Histogram of the nonzero contact forces in the cannonball packing.
  }
\end{figure}

\begin{figure}[H]
  \centering
  \begin{subfigure}{.4\textwidth}
    \includegraphics[width=\textwidth]{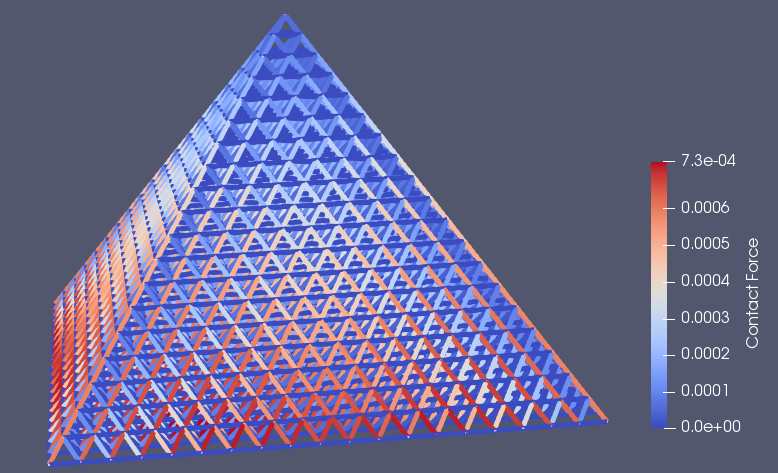}
    \caption{}
    \label{fig::CannonballZeroRestitutionHertzianCDMasonry}
  \end{subfigure}
  \begin{subfigure}{.4\textwidth}
    \includegraphics[width=\textwidth]{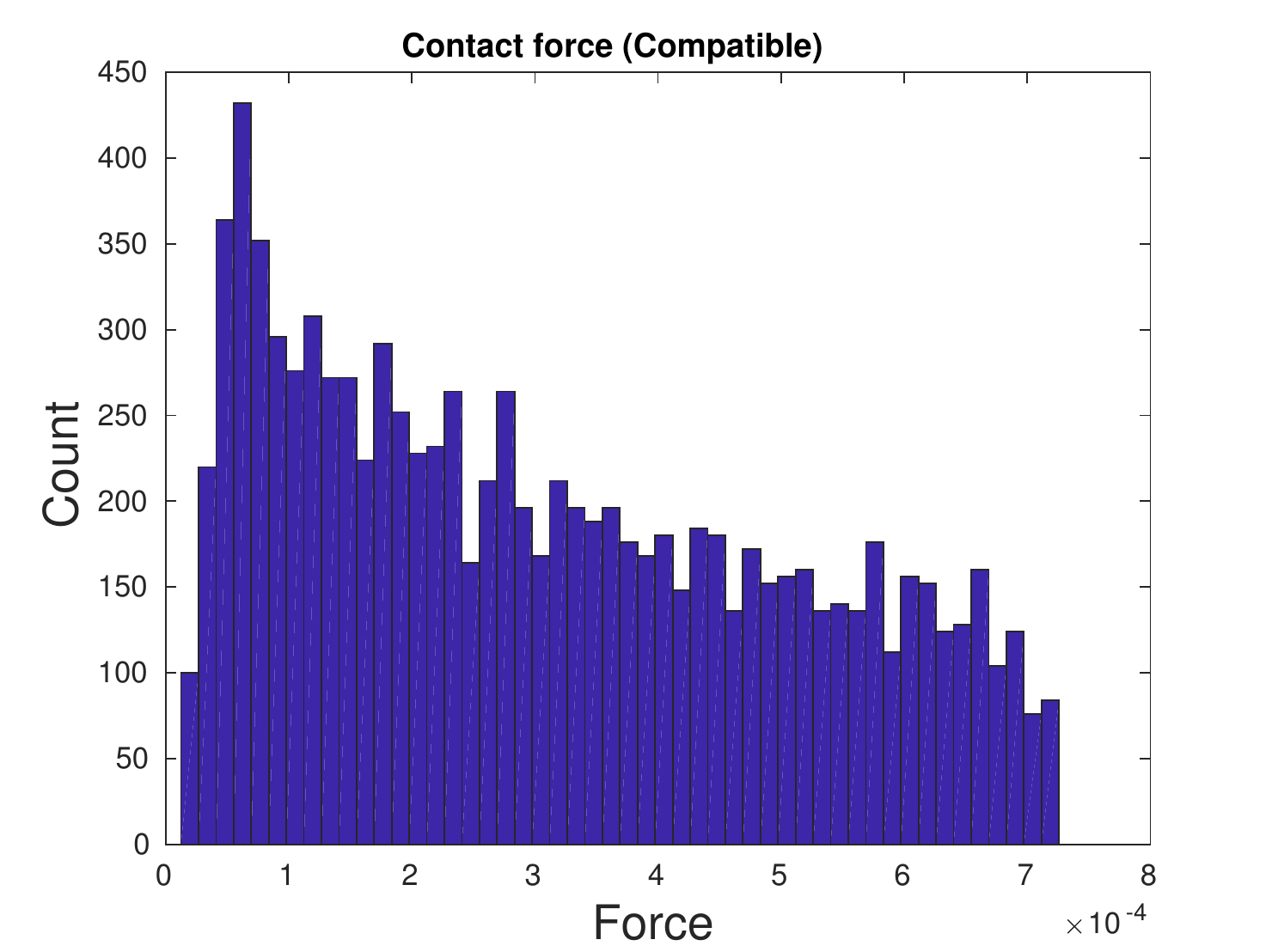}
    \caption{}
    \label{fig::CannonballZeroRestitutionHertzianCDMasonryHist}
  \end{subfigure}
  \caption{
    (a) The force distribution of compatible CD matches the reference force
    distribution computed by DEM in figure \ref{fig::CannonballZeroRestitutionHertzianDEM}.
    This plot was generated from a run wherein the classic CD problem was solved
    using the APGD algorithm, but the results are the same if PGS is used instead.
    (b) The histogram confirms the near-perfect match between the contact forces
    shown here and those in \ref{fig::CannonballZeroRestitutionHertzianDEMHist}.
    The quality of the match is determined by the tolerances to which the CD and
    compatiblility NLPs are solved.
  }
\end{figure}

\begin{figure}[H]
  \centering
  \includegraphics[width=.5\textwidth]{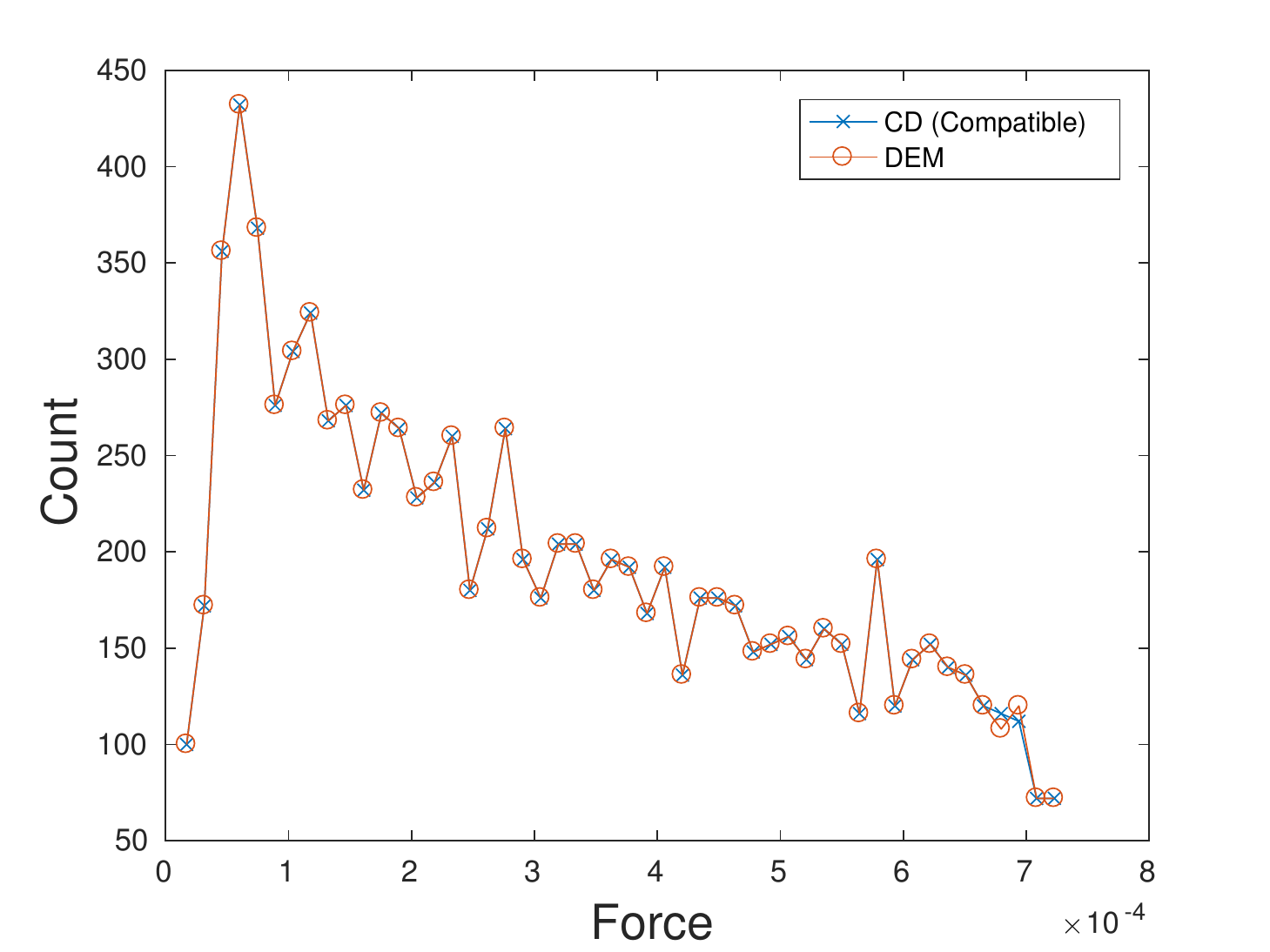}
  \caption{
    Contact force histograms from the DEM reference in figure \ref{fig::CannonballZeroRestitutionHertzianDEMHist}
    and compatible Hertizian CD in figure \ref{fig::CannonballZeroRestitutionHertzianCDMasonryHist}
    plotted side-by-side to show more clearly the close agreement between the two
    distributions. The small disagreement in some of the buckets is due to the
    finite solver tolerance in the CD and compatibility problems.
  }
  \label{fig::CombinedCannonballZeroRestitutionHertzianAPGDHist}
\end{figure}

\subsection{Floor Pressure}
\begin{figure}
  \centering
  \includegraphics[width=\textwidth]{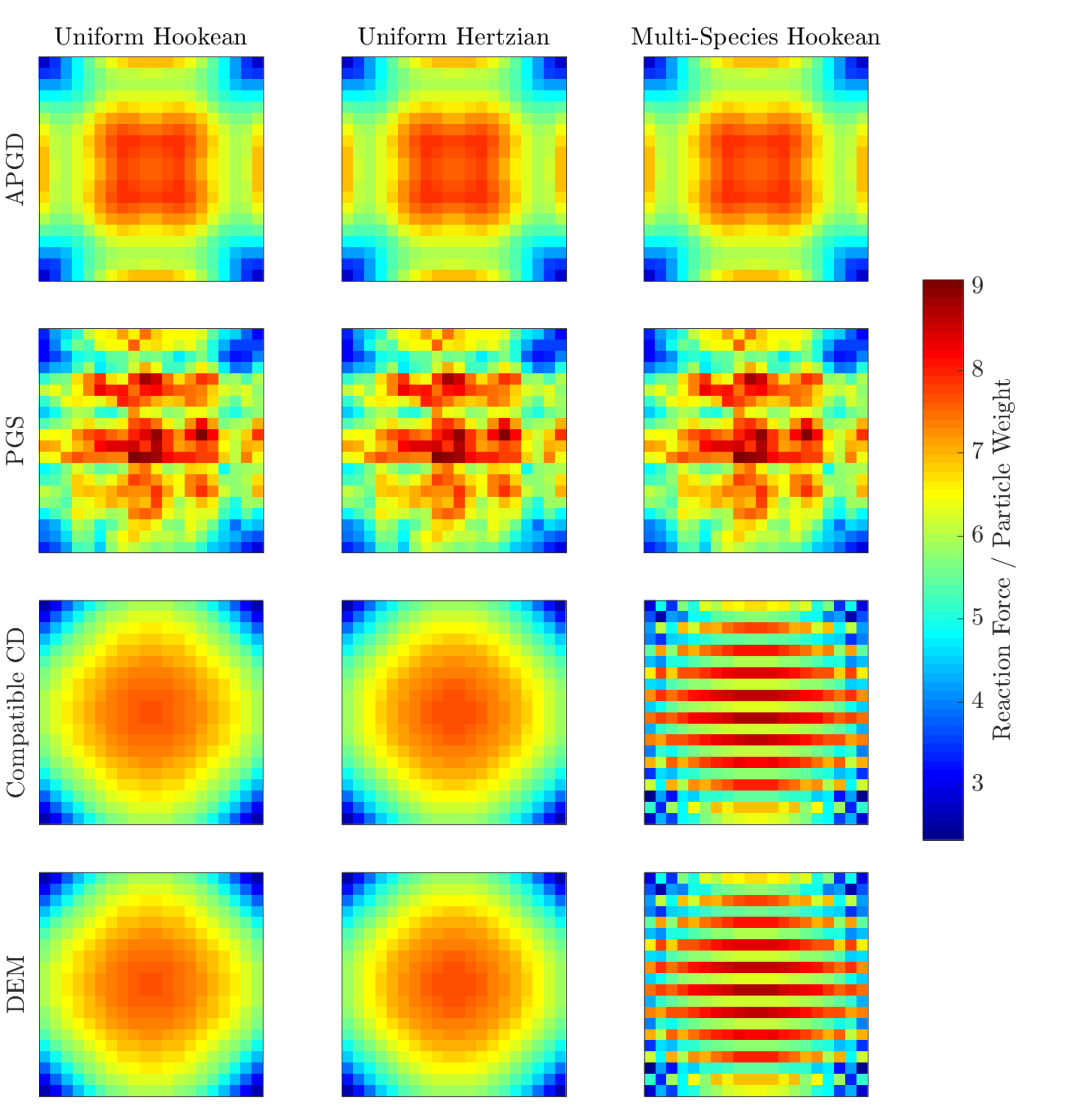}
  \caption{
    The contact model and solution algorithm has a great impact on the distribution
    of reaction forces on the base of the pyramid. Here, we show that the force distribution
    predicted by compatible CD matches the distribution computed by DEM. Each column
    of images corresponds to a particular contact model, denoted in the header.
    Each row corresponds to a single solution algorithm, denoted at left. All images
    use the same scale at right.
  }
  \label{fig::Pressure}
\end{figure}

Another measure of interest is the predicted distribution of reaction forces
beneath the base of the pyramid.
The goal here was to determine how the predicted force distribution changes
when you change the contact model and solution algorithm.
It can be seen in figure \ref{fig::Pressure} that the APGD and PGS algorithms
for CD do not change when the contact model changes, since they have no notion of
material behavior.
The Compatible CD solution method quantitatively matches the force distribution
predicted by the DEM method.
It should be emphasized that although the APGD solver produces reasonable-looking
contact forces, they do not match the forces computed by Compatible CD or DEM,
and in fact yield qualitatively different reaction force distributions.
As expected, the PGS solver produces a force distribution with no particular
order or symmetry.

The relative error in the pressure distribution predicted by Compatible CD is
shown for each case in table \ref{tab::PressureError}. The relative error
is computed by $\|F_{CD} - F_{DEM}\|/\|F_{DEM}\|$, where $F_{CD}$ is the
reaction force distribution supporting the base of the pyramid computed by
one of APGD, PGS, or Compatible CD algorithms.
We include the relative error of the APGD and PGS methods only to emphasize
the close match between Compatible CD and DEM.
The APGD and PGS methods produce qualitatively different results, so we do
not expect them to have small errors.

\begin{table}[H]
  \centering
  \begin{tabular}{c | c | c | c}
    Contact Model 
    & $\frac{\|F_{APGD} - F_{DEM}\|}{\|F_{DEM}\|}$
    & $\frac{\|F_{PGS} - F_{DEM}\|}{\|F_{DEM}\|}$
    & $\frac{\|F_{Compat} - F_{DEM}\|}{\|F_{DEM}\|}$\\
    \hline
    Uniform Hookean & 0.17 & 0.19 & $1.9 \cdot 10^{-5}$ \\
    Uniform Hertzian & 0.17 & 0.19 & $2.8 \cdot 10^{-5}$\\
    Multi-Species Hookean &  0.22 & 0.25 & $2.3 \cdot 10^{-5}$ 
  \end{tabular}
  \caption{Relative error in pressure distribution}
  \label{tab::PressureError}
\end{table}

\section{Discussion}
Contact dynamics is an efficient method for simulating the
interactions between rigid bodies, and is suitable for solving
very large systems over long periods of time.
Its main drawback, however, is the force indeterminacy that
arises as a result of the rigid body assumption.
We have detailed a method for recovering elastically-compatible
force solutions in the context of contact dynamics and
mention that the method can be easily integrated into existing
rigid-body codes due to the decoupled structure of the solution
algorithm.
The method can be applied to recover forces that would be
predicted by an overlap-penalty based discrete element method
in the limit of infinitely stiff particles.
We have applied the method to two contact models, linear (Hookean)
contacts and nonlinear (Hertzian) contacts, the two
most prominently-used contact models in discrete element
method literature.
The method is shown to provide contact forces numerically-identical
to those predicted by a high-stiffness penalty-based discrete element simulation
on a number of example problems where traditional CD is known to
fail due to indeterminacy.

We are currently developing a solution based on a numerical limiting
process of finite-stiffness elasticity contact dynamics, formulated in
detail in \cite{Krabbenhoft2012}.
In this method, the contact compliance is driven to zero
as the iterative method approaches an optimal point.
The appeal of this type of solution method is that it would
now solve the compatible CD problem in a single step, rather
than the two-step approach developed here.
It is a more computationally-efficient approach, since only
a single QP (or NLP) must be solved per time step.
In addition, a single-step solution method is desired for
extending this method to frictional contacts, where the
decoupling we have used is no longer possible due to the
non-unique velocity updates that arise when friction is considered.
This work is in development and will be published in a future article.

\newpage
\bibliographystyle{plain}
\bibliography{citations.bib}

\end{document}